\documentclass[prc,aps,float,twocolumn,superscriptaddress,showkeys,showpacs,nofootinbib]{revtex4}

\usepackage{amsmath}
\usepackage{amssymb}
\usepackage{amsfonts}
\usepackage{epsfig}
\usepackage{color}

\newcommand{\emb}{{\rule{0cm}{0cm}}}

\begin{document}

\title{Configuration Mixing within the Energy Density Functional Formalism: \\
Removing Spurious Contributions from Non-Diagonal Energy Kernels}

\author{D. Lacroix}
\email{lacroix@ganil.fr}
\affiliation{National Superconducting Cyclotron Laboratory, 1 Cyclotron Laboratory,
East-Lansing, MI 48824, USA}
\affiliation{GANIL, CEA et IN2P3, BP 5027, 14076 Caen Cedex, France}

\author{T. Duguet}
\email{thomas.duguet@cea.fr}
\affiliation{National Superconducting Cyclotron Laboratory,
1 Cyclotron Laboratory, East-Lansing, MI 48824, USA}
\affiliation{Department of Physics and Astronomy,
Michigan State University, East Lansing, MI 48824, USA}
\affiliation{CEA, Centre de Saclay, IRFU/Service de Physique Nucl{\'e}aire, F-91191 Gif-sur-Yvette, France}

\author{M. Bender}
\email{bender@cenbg.in2p3.fr}
\affiliation{Universit{\'e} Bordeaux,
             Centre d'Etudes Nucl{\'e}aires de Bordeaux Gradignan, UMR5797,
             F-33175 Gradignan, France}
\affiliation{CNRS/IN2P3,
             Centre d'Etudes Nucl{\'e}aires de Bordeaux Gradignan, UMR5797,
             F-33175 Gradignan, France}

\date{\today}



\begin{abstract}
Multi-reference calculations along the lines of the Generator Coordinate Method or the restoration of broken
symmetries within the nuclear Energy Density Functional (EDF) framework are becoming a standard tool in nuclear
structure physics. These calculations rely on the extension of a single-reference energy functional, of the Gogny
or the Skyrme types, to non-diagonal energy kernels. There is no rigorous constructive framework for this
extension so far. The commonly accepted way proceeds by formal analogy with the expressions obtained when applying
the generalized Wick theorem to the non-diagonal matrix element of a Hamilton operator between two product states.
It is pointed out that this procedure is ill-defined when extended to EDF calculations as the generalized Wick
theorem is taken outside of its range of applicability. In particular, such a procedure is responsible for the
appearance of spurious divergences and steps in multi-reference EDF energies, as was recently observed in
calculations restoring particle number or angular momentum. In the present work, we give a formal analysis of the
origin of this problem for calculations with and without pairing, i.e.\ constructing the density matrices from
either Slater determinants or quasi-particle vacua. We propose a method to regularize non-diagonal energy kernels such that
divergences and steps are removed from multi-reference EDF energies. Such a removal is a priori quasi-particle-basis dependent. A special feature of the method we use to proceed to the actual regularization is that it singles out one basis among all possible ones. The regularization method is applicable to calculations based on any symmetry restoration or generator coordinate but is limited to EDFs depending only on \emph{integer} powers of the normal and anomalous density matrices. Eventually, the method is formally illustrated for particle number restoration and is specified to configuration mixing calculations based on Slater determinants.
\end{abstract}

\maketitle

%
%

\section{Introduction}
\label{intro}

The nuclear Energy Density Functional (EDF) approach is the microscopic tool of choice to study medium-mass and
heavy nuclei in a systematic manner~\cite{bender03b}. The EDF approach used in nuclear physics has formal
similarities with Density Functional Theory (DFT)
\cite{hohenberg64,kohn65,kohn98,Nagy98DFT,dreizlerBook,parrBook,lecturenotesFNM} which provides a formal framework
to obtain the exact ground-state energy and one-body density of electronic many-body systems in condensed-matter
physics and quantum chemistry \cite{koch01}. However, and even if it is often referred to as \emph{nuclear DFT}
\cite{Bei74a,petkovBook,fayans,lecturenotesLGV,bulgac3,furnstahl05a}, the nuclear EDF method has deeply rooted
conceptual differences with standard Density Functional Theory; e.g.\ Ref.~\cite{Eng07a}.

One of the major challenges of the nuclear many-body problem is that atomic nuclei are self-bound strongly
correlated composite systems that tend to display strong collective modes and individual excitations on the same
energy scale. This makes the modeling of nuclei rather different from that of electronic systems in external
potentials to which DFT is heavily applied. The underlying assumption to nuclear EDF approaches is that
correlations can be divided into different classes that can be incorporated through successive refinements of the
method. While short range in-medium correlations are subsumed into a suitable energy density functional,
long-range correlations associated with collective modes must be incorporated more explicitly. On this basis, two
different levels of EDF calculations coexist in nuclear physics.

On the first level, inappropriately called self-consistent \emph{mean-field} theory, the EDF is constructed from a
one-body density matrix that corresponds to a single product state. This reference state is either taken as a
Slater determinant, abusively referred to as Hartree-Fock (HF) state, or as a quasi-particle product state,
abusively referred to as Hartree-Fock-Bogoliubov (HFB) state. In the present work, this level of description will
be referred to as the Single Reference Energy Density Functional (SR-EDF) method. Within the SR-EDF approach,
static correlations associated with collective modes are incorporated through the use of a symmetry-breaking
product state.

On the second level, several of those symmetry-breaking product states are mixed in the framework of the Generator
Coordinate Method (GCM) with the aim of restoring broken symmetries, allowing for fluctuations of the order
parameters of broken symmetries, or both. Such a Multi-Reference (MR) EDF approach allows one to incorporate
correlations associated with large amplitude collective motion beyond the static correlations which are easily
treated within the SR-EDF method. The energy functional at play in MR-EDF calculations depends on the transition
density matrices constructed from all possible pairs of product states that enter the MR set.

Until now, there exists neither a rigorous formal framework underlying MR-EDF calculations nor a guidance from
DFT. Indeed, the nuclear MR-EDF method is outside the scope of existing implementations of the Hohenberg-Kohn
theorem, even of those dealing specifically with symmetry breaking
issues~\cite{gross88a,gorling93a,Lei97aDFT,Har04aDFT}. As far as motivating MR-EDF calculations based on the GCM,
hybrid approaches used in electronic systems that combine a density functional for diagonal energy kernels and
(sometimes scaled) matrix elements of the bare Coulomb force for off-diagonal ones~\cite{Orestes07DFT} have to be
ruled out because of the different nature of the nuclear force and the more complex correlations it induces in the
nuclear medium. There is a consensus among practitioners that the SR-EDF should be recovered for any diagonal
energy kernel appearing in the MR-EDF and that the non-diagonal kernels should be extrapolated from the SR-EDF on
the basis of the Generalized Wick Theorem (GWT)~\cite{balian69a} which provides an efficient tool to evaluate the
matrix elements of any operator between two different product states. For diverse reasons, however, SR and MR-EDFs
do not reduce to matrix elements of a Hamiltonian operator between many-body states. As a
consequence, the procedure outlined above to generate the MR-EDF is not free from problems and inconsistencies.

The present article is the first paper of a trilogy were we address one of the problems arising from the current lack of a stringent and consistent constructive framework for the MR-EDF method, namely the appearance of divergences and finite steps that have been recently identified in the context of Particle Number Restoration (PNR)~\cite{donau98,almehed01a,anguiano01b,doba05a,flocard97}. We propose here a general cure for any type of configuration mixing. In Ref.~\cite{bender07x} the method is applied to PNR while in Ref.~\cite{duguet07a} the specific case of an EDF depending of non-integer powers of the density matrix is discussed. The spurious steps and divergences arise typically
as one scans the symmetry-restored energy as a function of a certain degree of freedom; e.g. looking at the PNR
potential energy surface of a nucleus as a function of axial quadrupole deformation. For reasons that will become
clear below, pure PNR is the only case where the divergent contributions to the energy functional can be traced
back rather easily although no practical solution to this problem has been proposed so far. It is important to
realize, however, that similar problems may appear for \emph{any} MR calculation performed within the EDF
framework, although they will be hidden in the formal expressions and may be masked by the numerical
representation. A pathology that resembles the one identified for PNR has recently been seen in EDF calculations,
without explicit treatment of pairing, aiming at restoring angular momentum~\cite{doba06a}, whereas a similar
problem was identified much earlier in the GCM-type mixing of zero- and two-quasi-particle
states~\cite{tajima92a}.

\begin{table}
\begin{center}
\begin{tabular}{ll}
\hline \hline \\
DFT &   Density Functional Theory \\
EDF &   Energy Density Functional \\
SWT &   Standard Wick Theorem \\
GWT &   Generalized Wick Theorem \\
SR  &   Single Reference \\
MR  &   Multi Reference \\
PNR &   Particle Number Restoration \\
PNP &   Particle Number Projection \\
BMZ &   Bloch-Messiah-Zumino \\
& \\
\hline \hline
\end{tabular}
\caption{List of acronyms repeatedly used in the text.}
\label{table:acronym}
\end{center}
\end{table}

It is essential to state that these difficulties are inherent to the EDF formulation of MR calculations and do not
exist when the energy relates to the strict average value of a Hamiltonian in a
wave-function \cite{donau98,almehed01a,anguiano01b}. As the major part of the paper consists of lengthy formal
manipulations that will be easier to follow if anticipating the results, we now summarize the three goals of the
present work:
\begin{itemize}
\item[(i)]
We want to introduce a formal framework that allows the unambiguous identification of terms in the EDF that are
responsible for divergences and finite steps in MR-EDF calculations, irrespective of the nature of the
configuration mixing involved. As will be shown below, finding the canonical basis of the Bogoliubov
transformation which links two HFB states allows us to do so. In that basis, the non-diagonal energy kernel
associated with those two vacua and which enters the MR energy can be constructed on the basis of the Standard
Wick theorem instead of the Generalized Wick Theorem. This will be the key to our analysis.
\item[(ii)] Guided by
the strict Hamiltonian case, several sources of problems including those responsible for divergences and steps are
unambiguously identified within the EDF context.

\item[(iii)]
We provide a minimum but general solution to the problem of divergences and steps in MR-EDF calculations for a
particular class of energy functionals. To that end, we propose a correction that removes exactly the difference
between the GWT-based and the SWT-based non-diagonal energy kernels. The latter correction preserves the
continuity condition between SR and MR-EDFs. Finally, we provide explicit expressions for particle-number
restoration and mixing of Slater determinants.
\end{itemize}

The paper is organized as follows: Section \ref{sect:why:mr:edf} briefly reviews general aspects of nuclear EDF
approaches. Section~\ref{basics} is devoted to the actual introduction to SR and MR-EDF methods. First, the
relevant similarities and key differences between Hamiltonian-based calculations and EDF ones are worked out.
Second, the connection between the SR functional and the non-diagonal energy kernels at play in the MR formalism
is discussed. In Section~\ref{illdefined}, we demonstrate that the usual definition of non-diagonal energy kernels
based on the GWT leads to ill-defined calculations when performed within the EDF approach. In
Section~\ref{correctingMREDF} we propose a correction removing the spurious contributions to the energy functional
that is applicable to any MR-EDF calculation. The correction is discussed in detail in Sections~\ref{PNR}
and \ref{sectionzeropairing}
for two applications of current interest. Conclusions and perspectives are given in Section~\ref{conclusions}.

%
%
\section{General aspects of the nuclear EDF approach}
\label{sect:why:mr:edf}

The nuclear energy density functional integrates out short-range correlations associated with non-collective
excitations on a given vacuum, in particular the strong tensor correlations from the vacuum nucleon-nucleon force,
thereby introducing higher-order density dependencies into the energy functional. The diagrams that are summed up
in the particle-hole and particle-particle channels are different~\cite{henley,klemt76a}, which strongly suggests the use
of different effective vertices in those two channels of the EDF. Lacking an explicit connection to first
principles, existing functionals are constructed using the symmetries of the nuclear Hamiltonian as a guiding
principle whereas their parameterizations are adjusted phenomenologically~\cite{bender03b}.

Allowing the reference state to break the symmetries of the eigenstates of the underlying Hamiltonian is a way to
incorporate \emph{static} long-range correlations associated with collective modes, as for example deformation and
pairing, with very moderate effort. However, the breaking of symmetries (translational, rotational, parity,
particle number, to name the most current ones) forbids a trivial connection of the nuclear SR-EDF formalism to
the Hohenberg-Kohn (HK) theorem which is the foundation of DFT \cite{Eng07a}. Indeed, the density that minimizes
the exact HK energy functional must reflect the symmetries of the exact ground state of the system. In fact, the
appearance of symmetry-breaking solutions in nuclear EDF calculations underlines two important elements (i) it is
crucial and numerically not too difficult to grasp the most important \emph{static} correlations using rather
simple approximate functionals and a single-determinantal reference state, at the price of violating the HK
theorem (ii) kinematical correlations associated with the corresponding symmetry modes (Goldstone modes) as well
as the correlations due to the fluctuation of their order parameters are extremely difficult to incorporate into a
single-determinantal approach. In other words, correlations associated with highly non-local processes such as
large-amplitude collective motions can hardly be described within a SR approach based on a standard, nearly-local
EDF.

It is possible to improve on the SR level by extending the EDF formalism to a Multi-Reference framework,
inappropriately referred to as \emph{beyond-mean-field} calculations in the literature. In a MR calculation, a set
of (usually) non-orthogonal product states of reference are combined to construct a more general energy functional
which involves the computation of non-diagonal norm overlaps and energy kernels between all pairs of vacua. Such
MR-EDF calculations are inspired by projection techniques and the Generator Coordinate Method (GCM) whose
formalisms are well documented for Hamiltonian-based calculations~\cite{Man75a,Won75a,ring80a,sheikh00}. Symmetry
restoration and GCM provide an efficient way to incorporate both collective and single-particle dynamics into a
coherent quantum-mechanical formulation. In particular, the GCM allows the inclusion of correlations associated
with the fluctuation of the order parameters of the broken symmetries. Group theoretical
considerations~\cite{duguet06b}, the Random-Phase-Approximation limit of GCM-type EDF
calculations~\cite{Robledo07a}, as well as the requirement of continuity between SR and MR functionals discussed
in the introduction can be used to constrain the form of the energy kernels to be used in a MR formalism. In
particular, it can be shown that the latter energy kernels have to be constructed exclusively from {\it
transition} normal and anomalous one-body density matrices, cf.\ Eqns.~(\ref{transrho}-\ref{transkappastar}) below,
between the two reference states involved. These transition density matrices, however, are at the origin of a
major difficulty that arises in existing implementations of MR EDF calculations.

It is noteworthy that the description of particular systems and phenomena in electronic systems also calls for
extensions of the Kohn-Sham implementation of HK-DFT to deal with symmetry-breaking
solutions~\cite{gross88a,gorling93a,Lei97aDFT,Har04aDFT}. The latter extensions are still connected to
{the}\emb\ HK theorem and differ from MR-EDF calculations performed in nuclear physics to restore broken
symmetries.

%
%
\section{From single-reference to multi-reference functional frameworks}
\label{basics}

We consider the nuclear Hamiltonian under the form
\begin{equation}
\label{eq:hamil}
H =   \sum_{ij} t_{ij} \, a^+_i \, a_j
    + \frac{1}{4} \sum_{ijkl} \bar{v}_{ijkl} \, a^+_i \, a^+_j \, a_l \, a_k
\, ,
\end{equation}
which, for the sake of transparency of the chain of arguments below, has been limited to the sum of the kinetic
energy and a two-body interaction $v_{12}$, represented through its antisymmetrized matrix elements
$\bar{v}_{ijkl}$. For the moment we also do not distinguish between protons and neutrons. The extension to the
case of two nucleon species and three-body or even multi-body forces of higher order is straightforward (although
cumbersome) and will not influence our conclusions.

It is important to stress at this point that, in the present paper, $H$ defined above strictly refers to a
"vacuum" nuclear Hamiltonian operator; e.g. its two-body part reproduces nucleon-nucleon scattering data up to
relevant laboratory energies. Later in this section we refer to \emph{effective} Hamiltonians, either in the
context of a truncated single-particle space or because they re-sum correlations through an explicit dependence on
the density of the system. It happens that replacing the vacuum Hamiltonian $H$ by a genuine operator defined
within a truncated single-particle space would not modify the following discussion regarding the technical issue
addressed in the present paper, whereas using a density-dependent "Hamiltonian" is at the origin of some of the problems
we are concerned with.

%
\subsection{Single-reference framework}
\label{sec:discussion_edfwick}
In the SR approach, the reference state is taken to be an independent-quasi-particle state $| \Phi_0 \rangle$,
most conveniently written as a quasi-particle vacuum
\begin{equation}
\label{initialstate0}
| \Phi_0 \rangle
= \mathcal{C}_0 \, \prod_\nu \alpha_\nu | 0 \rangle \, ,
\end{equation}
where $\mathcal{C}_0$ is a complex normalization coefficient. The set of quasi-particle operators $\{\alpha_{\nu} ,
\alpha^{+}_{\nu}\}$ is obtained from the complete basis of single-particle operators $\{a_i, a^+_i\}$ through a
Bogoliubov transformation characterized by the matrices $(U^0,V^0)$
\begin{equation}
\label{eq:bogo}
\alpha^+_\nu
= \sum_i \big( U^0_{i \nu} \, a^+_i + V^0_{i \nu} \, a_i \big) \, .
\end{equation}
The index $"0"$ appearing in $| \Phi_0 \rangle$ as well as in $U^0$ and $V^0$ denotes an ensemble of collective coordinates
that fully characterizes the product state $| \Phi_0 \rangle$, either regarding broken symmetries or the average
value of a constraining operator. This label is of no explicit importance at the SR level but will become
mandatory for MR calculations.
%
%
\subsubsection{Strict HFB approach}

In the strict HFB approach \cite{Valatin61a}, the reference state $| \Phi_0 \rangle$ approximates the actual
ground-state of the system in such a way that the approximate energy is obtained through
\begin{eqnarray}
\label{eq:hamil_phi00}
E_0
& = & \frac{\langle \Phi_{0} | H | \Phi_{0} \rangle}
           {\langle \Phi_{0} | \Phi_{0} \rangle}
\, .
\end{eqnarray}
Using the Standard Wick theorem (SWT) \cite{wick50a}, Eq.~(\ref{eq:hamil_phi00}) can be expressed as
\begin{eqnarray}
E_0
& = & \sum_{ij} t_{ij} \, \rho^{00}_{ji}
      + \frac{1}{2} \sum_{ijkl} \bar{v}_{ijkl}\, \rho^{00}_{ki} \, \rho^{00}_{lj}
      \nonumber \\
&   & \phantom{ \sum_{ij} t_{ij} \, \rho^{00}_{ji} }
      + \frac{1}{4} \sum_{ijkl} \bar{v}_{ijkl} \, \kappa^{00 \, \ast}_{ij} \, \kappa^{00}_{kl}
\, , \label{eq:hamil_phi0}
\end{eqnarray}
where $\rho^{00}$ is the \emph{intrinsic} (SR) normal one-body density matrix and $\kappa^{00}$ is the intrinsic
anomalous density matrix (also called pairing tensor)
\begin{alignat}{2}
\rho^{00}_{ij}
& \equiv  \frac{\langle \Phi_{0} | a^{+}_{j} a_{i} | \Phi_{0} \rangle}
           {\langle \Phi_{0}  | \Phi_{0}  \rangle} &
& =   \left(V^{0 \, \ast} V^{0 \, T} \right)_{ij} \, ,
\label{intrrho}    \\
\kappa^{00}_{ij} & \equiv  \frac{\langle \Phi_{0} | a_{j} a_{i} | \Phi_{0} \rangle}
           {\langle \Phi_{0} | \Phi_{0} \rangle} &
& =   \left(V^{0 \, \ast} U^{0 \, T} \right)_{ij}
\label{intrkappa}  \, ,   \\
\kappa^{00 \, \ast}_{ij} & \equiv  \frac{\langle \Phi_{0} | a^{+}_{i} a^{+}_{j} | \Phi_{0} \rangle}
           {\langle \Phi_{0} | \Phi_{0} \rangle} &
& = \left(V^{0} U^{0 \, +} \right)_{ij} \label{intrkappastar}  \, ,
\end{alignat}
associated with the reference state $| \Phi_{0} \rangle$. The normal density matrix is hermitian $\rho^{00}_{ij} =
\rho^{00 \, \ast}_{ji}$, whereas the anomalous density matrix is skew symmetric $\kappa^{00}_{ij} =
-\kappa^{00}_{ji}$.

The energy $E_0$ given by Eq.~(\ref{eq:hamil_phi0}) can be seen as the particular energy functional $E
\left[\rho^{00}, \kappa^{00}, \kappa^{00 \, \ast} \right]$ that results from taking the expectation value of the
Hamiltonian in the product state. This is what is referred to as the \emph{strict} HFB approximation. The
minimization of $E \left[\rho^{00}, \kappa^{00}, \kappa^{00 \, \ast} \right]$ with respect to all independent
degrees of freedom ($\rho^{00}_{ii}$ as well as $\rho^{00}_{ij}$, $\rho^{00 \, \ast}_{ij}$, $\kappa^{00}_{ij}$,
and $\kappa^{00 \, \ast}_{ij}$ for $j < i$), under the constraints that the reference state $| \Phi_{0} \rangle$
remains a quasi-particle vacuum and that the average particle number $\langle \Phi_{0} | \hat{N} | \Phi_{0}
\rangle / \langle \Phi_{0} | \Phi_{0} \rangle$ has the fixed value $N_0$, leads to the HFB equation that
determines ($U^0, V^0$) and the densities.

So far, we have not discussed the characteristics of the nuclear Hamiltonian $H$. Traditional nucleon-nucleon
hard-core potentials are not perturbative~\cite{bethe71a}, which makes any attempt to use the strict HFB
approximation to the exact nuclear many-body problem useless. The recently proposed soft-core interactions on the
other hand seems to make the nuclear many-body problem perturbative~\cite{bogner05a}. Still, one must go beyond
lowest order to obtain close to converged results and the strict HFB approximation is not quantitatively viable.
The situation is further complicated by the necessity to treat three-body and maybe higher-body forces in the
nuclear many-body problem~\cite{grange89a,pieper01a,zuo02b,kuros02a,bogner05a}. As a consequence, one often
resorts to an \emph{effective Hamiltonian} that implicitly incorporates correlations brought by the physics
outside the model space used in a given calculation. Traditionally, there are two major classes of effective
interactions for self-consistent mean-field calculations that represent two different philosophies and strategies
to reduce the nuclear many-body problem to a tractable number of relevant degrees of freedom. There are (i)
approaches that can be seen as lowest-order approximations to the interacting shell model \cite{Cau05a},
formulated in terms of effective (often schematic) Hamiltonians in a strict HFB approach using a schematic
shell-model space of a very few spherical harmonic-oscillator $j$ shells \cite{Man75a} (ii) approaches that use
the full model space of occupied particles together with a density-dependent effective interaction, which are what
is nowadays recognized as an approximation to a more general SR-EDF formalism.

%
%
\subsubsection{EDF approach}

The EDF approach builds upon the last comment. In the context of an EDF formalism, the reference product state $|
\Phi_{0} \rangle$ used to construct the one-body normal and anomalous density matrices
(\ref{intrrho}-\ref{intrkappastar}) is to be seen as an \emph{auxiliary} state which is not meant to be a good
approximation of the actual ground-state wave function, in a similar manner as in DFT for electronic systems
\cite{kohn64a,kohn98,Nagy98DFT,dreizlerBook,parrBook,lecturenotesFNM,koch01}. The EDF and the HFB-like equation
which results from its minimization contain higher-order correlations than those provided by the strict Hartree,
Fock and Bogoliubov diagrams written in terms of the vacuum interaction. It is important to note that, as opposed to
the strict HFB method, the SR-EDF approach does not rely on the Ritz variational principle in the sense that the
minimal energy obtained from an approximate energy functional may be below the actual ground-state energy of the
system.

Thus, instead of the expectation value of $H$, the starting point of the method is an energy functional $\mathcal{E}
\left[\rho^{00}, \kappa^{00}, \kappa^{00 \, \ast} \right]$ of the normal and anomalous density matrices.
Considering the simple case of a bilinear functional in order to stay formally close to the strict HFB approach
with a two-body interaction, a typical EDF can be written as
\begin{eqnarray}
\label{eq:e00}
\mathcal{E}[\rho^{00}, \kappa^{00}, \kappa^{00 \, \ast}]
& \equiv & \mathcal{E}^{\rho}
           + \mathcal{E}^{\rho\rho}
           + \mathcal{E}^{\kappa\kappa}
           \nonumber \\
& = & \sum_{ij} t_{ij} \, \rho^{00}_{ji}
      + \tfrac{1}{2} \sum_{ijkl} \bar{v}^{\rho\rho}_{ijkl} \,
        \rho^{00}_{ki} \, \rho^{00}_{lj}
      \nonumber \\
&   & \phantom{ \sum_{ij} t_{ij} \, \rho^{00}_{ji}   }
      + \tfrac{1}{4} \sum_{ijkl} \bar{v}^{\kappa\kappa}_{ijkl} \,
        \kappa^{00 \, \ast}_{ij} \, \kappa^{00}_{kl}
\, .
\end{eqnarray}
In Eq.~(\ref{eq:e00}), the matrix elements of the effective vertex $\bar{v}^{\rho\rho}$ might or might not be
antisymmetric; e.g. due to the Slater approximation used to treat the exchange contribution from Coulomb or by
dropping specific terms of the Skyrme functional~\cite{bender03b}. Due to the antisymmetry of $\kappa$, however,
only the antisymmetrized part of the vertex is probed in the last term of Eq.~(\ref{eq:e00}) and one can always take
$\bar{v}^{\kappa\kappa}$ to be antisymmetric, which we will do here. In any case, and even though
Eqs.~(\ref{eq:hamil_phi0}) and~(\ref{eq:e00}) look very similar, $\bar{v}^{\rho\rho}$ and $\bar{v}^{\kappa\kappa}$
should not be seen as interactions in the real sense and are likely to be different. Of course, they are in
principle related to the original vacuum interaction but, in the absence of a constructive framework, the link
remains implicit. Popular energy density functionals for calculations along these lines~\cite{bender03b} are the
non-relativistic Fayans~ \cite{fayans}, Gogny~\cite{decharge80a} and Skyrme~\cite{vautherin72a} as well as the relativistic~~\cite{Vretenar05a} ones.

\subsection{Multi-reference extension}

Because the nuclear Hamiltonian is invariant under certain symmetry transformations, the independent
quasi-particle state $| \Phi_{0} \rangle$ should be an eigenstate of appropriate linear combinations of the infinitesimal generators.
However, the simultaneous preservation of symmetries and the inclusion of long-range correlations associated with
large-amplitude collective motions is impossible within the strict HFB formalism and remains untractable within a
SR-EDF formalism. As a result, the product state $| \Phi_{0} \rangle$ is permitted to spontaneously break the
symmetries of the true eigenstates to lead to energetically favored solutions; an ambiguity of mean-field and EDF
approaches for which L{\"o}wdin coined the notion of "symmetry dilemma" in Ref.~\cite{LowdinSymDil}.

To solve that ambiguity, one resorts to a multi-reference extension of the method~\cite{bender03b}. The starting
point consists of a finite set of product states $\{| \Phi_{0} \rangle; 0 \in \text{MR}\}$, where $\{0 \in \text{MR}\}$ denotes
different realizations of the set of collective coordinates that characterize a reference state. If focusing on
symmetry restoration, the states belonging to the MR set correspond to one another by the application of a
transformation of the symmetry group under consideration. However, the label of the product states in the MR set
can also refer to different values of the order parameter of the broken symmetry; the goal of mixing such states
being to incorporate correlations associated with fluctuations of that order parameter. In such a case, the
operator that links two vacua in the set is not known analytically.

\subsubsection{Strict projected-GCM approach}

In the strict projection-GCM approach, one constructs the projected-GCM state from the product states in the MR
set through
\begin{equation}
\label{MRwf}
| \Psi^{k} \rangle
= \sum_{\{0\} \in {\rm M\!R}} f^{k}_{0} \, | \Phi_{0} \rangle \, ,
\end{equation}
where the determination of the weight functions $f^{k}_{0}$ is discussed below. The index $k$ denotes that not
only the ground state, but also excited states can be extracted, either by projection, by diagonalization, or
both. In the \emph{strict} projection-GCM approximation, the many-body energy is approximated by the average value
of the nuclear Hamiltonian in the projected-GCM wave-function
\begin{eqnarray}
\label{MRenergystrict}
E^{k} & \equiv & \frac{\langle \Psi^{k} | \hat H | \Psi^{k} \rangle}
           {\langle \Psi^{k} | \Psi^{k} \rangle}
       \\
& = & \frac{\sum_{\{0,1\}\in {\rm M\!R}} f^{k \, \ast}_{0} \, f^{k}_{1} \, E[0,1] \,
            \langle \Phi_{0} | \Phi_{1} \rangle}
           {\sum_{\{0,1\}\in {\rm M\!R}} f^{k \, \ast}_{0} \, f^{k}_{1}\,
            \langle \Phi_{0} | \Phi_{1} \rangle}
\, , \label{MRenergystrict2}
\end{eqnarray}
where the set of energy $E[0,1] \equiv \langle \Phi_0 | \hat{H} | \Phi_1 \rangle/\langle \Phi_0 | \Phi_1 \rangle$
and norm $\langle \Phi_{0} | \Phi_{1} \rangle$ kernels constitute the basic ingredients of the method. Note that
Eq.~(\ref{MRenergystrict2}) is nothing but the energy in the (Hamiltonian-based) Generator Coordinate Method
(GCM)~\cite{Won75a,ring80a}. From a formal point of view, symmetry restoration is a special case of the GCM.
Indeed, in that case $| \Psi^{k}\rangle$ can be expressed as a projector acting on one of the symmetry-breaking
product state in the MR set. Whereas in the GCM the weight functions $f^{k}$ are determined variationally from the
Hill-Wheeler-Griffin equation \cite{Hill53,Griffin57}, they are dictated by properties of the relevant symmetry
group when performing symmetry restoration\footnote{The weight functions are fully determined only if the symmetry
group is abelian.}.

The energy kernel $E[0,1]$ can be evaluated through the use of the Generalized Wick Theorem (GWT)~\cite{balian69a}
\begin{eqnarray}
 E_{GWT}[0,1] & = & \sum_{ij} t_{ij} \, \rho^{01}_{ji}
      + \frac{1}{2} \sum_{ijkl} \bar{v}_{ijkl}\, \rho^{01}_{ki} \, \rho^{01}_{lj}
      \nonumber \\
&   & \phantom{ \sum_{ij} t_{ij} \, \rho^{01}_{ji}  }
      + \frac{1}{4} \sum_{ijkl} \bar{v}_{ijkl} \, \kappa^{10*}_{ij} \, \kappa^{01}_{kl}
\, . \label{eq:hamil_phi01}
\end{eqnarray}
It is essential for the present work to note that Eq.~(\ref{eq:hamil_phi01}) is exactly of the same functional form
as Eq.~(\ref{eq:hamil_phi0}), except that \emph{intrinsic} density matrices $\{\rho^{00}, \kappa^{00}, \kappa^{00 \,
\ast}\}$ have been replaced by \emph{transition} density matrices defined as
\begin{eqnarray}
\label{transrho}
\rho_{ij}^{01}
& \equiv & \frac{\langle \Phi_{0} | a^{+}_{j} a_{i}| \Phi_{1} \rangle}
                {\langle \Phi_{0}  | \Phi_{1}  \rangle} \, ,
    \\
\label{transkappa}
\kappa_{ij}^{01}
& \equiv & \frac{\langle \Phi_{0} | a_{j} a_{i}| \Phi_{1} \rangle}
                {\langle \Phi_{0} | \Phi_{1} \rangle} \, ,
      \\
\label{transkappastar}
\kappa^{10 \, \ast}_{ij}
& \equiv & \frac{\langle \Phi_{0} | a^{+}_{i} a^{+}_{j}| \Phi_{1} \rangle}
                {\langle \Phi_{0} | \Phi_{1} \rangle} \, ,
\end{eqnarray}
where $\kappa^{10 \, \ast}_{ij}$ is in general not the complex conjugate of $\kappa_{ij}^{01}$ anymore. Given
Eq.~(\ref{eq:hamil_phi01}), it is clear that any "diagonal" energy kernel, i.e.\ obtained for $| \Phi_1 \rangle = |
\Phi_0 \rangle$, is equal to the strict HFB energy, Eq.~(\ref{eq:hamil_phi0}), so that the continuity requirement
between the SR energy and MR energy kernels stated in the introduction is trivially fulfilled.

As in the strict HFB approximation, a strict projected-GCM calculation performed in terms of the vacuum nuclear
Hamiltonian does not lead to quantitatively satisfactory results for all observables of interest. Still, the
restoration of the Galilean invariance of spherical HF states~\cite{rodriguez04a,rodriguez04b} was studied
within such a scheme; i.e.\ omitting the density-dependent part of the Gogny interaction and keeping exactly
all exchange terms in the computation of MR kernels. In fact, MR calculations employing Hamiltonian operators
are rather performed within a limited valence space, in the spirit of the interacting shell
model, either of projected-GCM type \cite{bur95,Ena99a,Ena01a,Ena02a} or using more elaborate schemes to select
the set of reference states like in the MONSTER and VAMPIR approaches \cite{Schm87a,schmid04,Don92a}.

\subsubsection{EDF approach}

To perform quantitatively relevant calculations using the full model space of single-particle states, one has to
turn to an energy density functional variant of MR calculations. Guided by Eq.~(\ref{MRenergystrict2}), the
many-body energy is defined in the MR-EDF approach by
\begin{equation}
\mathcal{E}^{k}
\equiv \frac{\sum_{\{0,1\} \in {\rm M\!R}} f^{k \, \ast}_{0} \, f^{k}_{1} \, \mathcal{E}[0,1] \,
        \langle \Phi_{0} | \Phi_{1} \rangle}
       {\sum_{\{0,1\} \in {\rm M\!R}} f^{k \, \ast}_{0} \, f^{k}_{1} \,
        \langle \Phi_{0} | \Phi_{1} \rangle}
\, , \label{MRenergy}
\end{equation}
which now depends only implicitly on the projected-GCM state of Eq.~(\ref{MRwf}). Indeed, the energy
$\mathcal{E}^{k}$ is not the average value of $H$, or any genuine operator\footnote{A density-dependent operator,
that is an operator that depends on the solution, is not considered in the present work as a \emph{genuine}
operator.}, in $| \Psi^{k} \rangle$ and it cannot be re-expressed into Eq.~(\ref{MRenergystrict}). Rather,
$\mathcal{E}^{k}$ is to be seen as a more general functional of all product states belonging to the MR set $\{|
\Phi_{0} \rangle ; 0 \in MR\}$ as each energy kernel $\mathcal{E}[0,1]$ is a functional of a pair of states $\{|
\Phi_{0} \rangle ; | \Phi_{1} \rangle\}$ in the set.

Although Eq.~(\ref{MRenergy}) provides the general ground to calculate the energy of the system within the MR-EDF
formalism, there remains the question of how the energy kernel $\mathcal{E}[0,1]$ should be constructed and what its
connection to the SR energy functional $\mathcal{E}[\rho^{00}, \kappa^{00}, \kappa^{00 \, \ast}]$ introduced in the
previous section is. Guided by the structure of the expressions in the strict HFB and GCM theories,
Eqs.~(\ref{eq:hamil_phi0}) and~(\ref{eq:hamil_phi01}), practitioners of nuclear EDF methods
\cite{anguiano01b,Rei94a,Rod02a,Robledo07a,bonche90,heenen93a} have used the natural extension $\mathcal{E}[0,1] \equiv
\mathcal{E}[\rho^{01}, \kappa^{01}, \kappa^{10 \, \ast}]$ to define the MR energy kernel entering Eq.~(\ref{MRenergy}) from
the SR energy functional. Using the bilinear SR functional of Eq.~(\ref{eq:e00}) for illustration, this leads to a
kernel of the form
\begin{eqnarray}
\label{eq:e01} \mathcal{E}_{GWT}[0,1] & \equiv & \sum_{ij} t_{ij} \, \rho^{01}_{ji}
      + \tfrac{1}{2} \sum_{ijkl} \bar{v}^{\rho\rho}_{ijkl} \,
        \rho^{01}_{ki} \, \rho^{01}_{lj}
      \nonumber \\
&   & \phantom{ \sum_{ij} t_{ij} \, \rho^{01}_{ji}  }
      + \tfrac{1}{4} \sum_{ijkl} \bar{v}^{\kappa\kappa}_{ijkl} \,
        \kappa^{01 \, \ast}_{ij} \, \kappa^{10}_{kl}
\, ,
\end{eqnarray}
which is labeled as "GWT" since the definition $\mathcal{E}[0,1] \equiv \mathcal{E}[\rho^{01}, \kappa^{01},
\kappa^{10 \, \ast}]$ amounts to using the generalized Wick theorem as a motivation to define the MR energy kernel
from the SR-EDF.

Calculations along these lines using realistic EDFs have been performed for the restoration of particle
numbers~\cite{heenen93a,anguiano01b} and angular momentum of axially-deformed HFB
states~\cite{bender03c,duguet03c,bender04a,egido04a}, of angular momentum of cranked HF
states~\cite{baye84a,doba06a}, and of parity of octupole deformed HFB states~\cite{egido91a,heenen94a}.
Some of those calculations also included a GCM-type configuration mixing along a collective degree
of freedom related to spatial deformation.

Other prescriptions than the simple replacement of $\rho^{00}_{ki}$ with $\rho^{01}_{ki}$ have been tried but were
found to lead to unrealistic results~\cite{Robledo07a}. The prescription $\mathcal{E}[0,1] \equiv \mathcal{E}[\rho^{01},
\kappa^{01}, \kappa^{10 \, \ast}]$ ensures that the continuity requirement between the SR and MR levels of
description is fulfilled. For symmetry restorations, it can also be shown that the functional should only depend
on transition densities~\cite{duguet06b}.

However, the GWT-based prescription to define $\mathcal{E}[0,1]$ causes serious problems as was recently illustrated
for PNR in Refs.~\cite{anguiano01b,doba05a} and confirmed from a different point of view in Ref.~\cite{bender07x},
after earlier warnings~\cite{tajima92a,donau98,almehed01a} were not recognized. The most spectacular manifestation
relates to the appearance of divergences in the PNR energy when two vacua in the MR set are orthogonal,
which in pure PNR applications happens for a relative gauge angle of $\pi/2$ when a pair of single-particle states
has the occupation probability of $1/2$~\cite{anguiano01b}. While this does not pose any problem in the strict
PNP-HFB method~\cite{donau98,almehed01a}, this makes PNR to be ill-defined within the MR-EDF approach. In fact,
the energy functional does not only contain divergences when encountering orthogonal vacua, but also finite
spurious contributions even when the two vacua are not orthogonal~\cite{bender07x} as will be illustrated later in
the present work. It is the goal of the following sections to introduce a method that allows the identification
and removal of these spurious terms from any type of MR-EDF calculations.

It is noteworthy that, to the best of our knowledge, the continuity requirement that is always used as a guiding
principle to construct nuclear MR energy kernels is never enforced in MR calculations for electronic systems.
There, hybrid approaches that use a density functional for diagonal kernels and the (sometimes scaled) bare
Coulomb force for non-diagonal kernels are used instead~\cite{Lei97aDFT,Gri96aDFT,Gri99aDFT,Har04aDFT}, an
approach that is facilitated by the fact that the correlation and exchange parts of the energy are a mere
correction to the direct Coulomb interaction in most Coulomb systems.

%
%
\section{Energy kernels in MR approaches}
\label{illdefined}

Although the main point of the present work is to stress the differences between strict HFB/projected-GCM
approaches and EDF ones, the analysis of the former does offer the key to the understanding of the problems
encountered in the latter. As a result, we first concentrate on the strict projected-GCM method and compare
the computation of the energy kernel $\langle \Phi_0 | H | \Phi_1 \rangle/\langle \Phi_0 | \Phi_1 \rangle$ as
obtained from both the GWT and the SWT in a suitable basis.

\subsection{Notation and preliminary discussion}

In this section we summarize the elements which will be needed below for the analysis of MR-EDF calculations. The
key ingredient will be to find a (quasi-particle) basis that allows the computation of the energy kernel $\langle \Phi_0 |
H | \Phi_1 \rangle/\langle \Phi_0 | \Phi_1 \rangle$ in terms of the SWT instead of the commonly used GWT. The
question if and how this is possible has been addressed already from various perspectives in the
past~\cite{ring80a,nee83,bur95,Dob00}.
\begin{enumerate}
\item We write the two vacua involved as products of the two corresponding (different) sets of quasi-particle
operators, denoted by $\alpha_\nu$ and $\beta_\nu$, respectively,
\begin{eqnarray}
\label{initialstates}
| \Phi_0 \rangle
& = & \mathcal{C}_0 \, \prod_{\nu} \alpha_\nu | 0 \rangle
       \, , \nonumber \\
| \Phi_1 \rangle
& = & \mathcal{C}_1 \, \prod_{\mu} \beta_\mu | 0 \rangle \, .
\end{eqnarray}
The moduli of the two complex constants $\mathcal{C}_0$ and $\mathcal{C}_1$ are fixed by the normalization of the
states, while their phases can be set to any value.
\item
We introduce two sets of matrices $(U^0,V^0)$ and $(U^1,V^1)$ that represent the Bogoliubov transformations
between an arbitrary, but common, single-particle basis $\{a_i, a^+_i\}$ and the two sets of quasi-particle states
\begin{eqnarray}
\label{eq:bogo0}
\alpha^+_\nu
& \equiv & \sum_i \big( U^0_{i \nu} \, a^+_i + V^0_{i \nu} \, a_i \big) \, ,
      \\
\beta^+_\nu & \equiv & \sum_i \big( U^1_{i \nu} \, a^+_i + V^1_{i \nu} \, a_i \big) \, .
\end{eqnarray}
\item
One realizes that the transformation that expresses the quasi-particle operators $\{\beta_{\mu},
\beta^+_{\mu}\}$ associated with $| \Phi_1 \rangle$ in terms of the quasi-particle operators $\{\alpha_{\nu},
\alpha^+_\nu\}$ associated with $| \Phi_0 \rangle$ is \emph{canonical}, which means it has the form of a general
Bogoliubov transformation
\begin{equation}
\label{eq:uv1}
\beta^+_\mu
= \sum_{\nu}        \big(   A_{\nu\mu} \, \alpha^+_{\nu}
                          + B_{\nu\mu} \, \alpha_{\nu}
                    \big) \, ,
\end{equation}
with the matrices $A$ and $B$ satisfying (see Appendix~E of Ref.~\cite{ring80a})
\begin{eqnarray}
\label{eq:uv2} A
& \equiv & U^{0^+} U^1 + V^{0^+} V^1 \,,
        \\
B
& \equiv & V^{0^T} U^1 + U^{0^T} V^1 \, . \label{eq:uv3}
\end{eqnarray}
To avoid any misunderstanding in what follows, we stress that the Bogoliubov transformation~(\ref{eq:uv1})
represented by the $A$ and $B$ matrices has nothing to do with pairing correlations. As a consequence, its
ingredients have a particular meaning and interpretation that might seem unusual and unexpected (e.g.\ see
Sec.~\ref{sectionzeropairing}).

We also stress that no restriction or symmetry is presently imposed on the reference states $\{| \Phi_{0} \rangle
; 0 \in \text{MR}\}$. They might break time-reversal invariance and have an odd number-parity~\cite{ring70a,banerjee73a} (describing
odd nuclei), be two-quasi-particle states (describing diabatic excitations), or higher
quasi-particle states. The only restriction is that all reference states belonging to the MR set have the same definite number parity.
\item
The matrices $A$ and $B$ defined through Eqs.~(\ref{eq:uv1}-\ref{eq:uv3}) can be decomposed into a sequence of
three simpler transformations thanks to the Bloch-Messiah-Zumino (BMZ) theorem~\cite{bloch,zumino}
\begin{equation}
\label{eq:cano}
A \equiv D \bar{A} C  \, , \qquad B \equiv D^* \bar{B} C \, .
\end{equation}
where $D$ ($C$) is a unitary transformation which only mixes quasi-particle creation/annihilation operators
$\{\alpha^+_{\nu}\}/\{\alpha_{\nu}\}$ ($\{\beta^+_{\mu}\}/\{\beta_{\mu}\}$) among each other, whereas $\bar{A}$
and $\bar{B}$ represent a special Bogoliubov transformation. The transformations $D$ and $C$ of the decomposition (\ref{eq:cano}) introduces two
intermediate quasi-particle bases $\{\tilde{\alpha}_{\nu}, \tilde{\alpha}^+_{\nu}\}$ and $\{\tilde{\beta}_{\mu},
\tilde{\beta}^+_{\mu}\}$ with different properties from the two original ones $\{\alpha_{\nu}, \alpha^+_{\nu}\}$
and $\{\beta_{\mu}, \beta^+_{\mu}\}$. As a quasi-particle vacuum is invariant under a unitary transformation among
quasi-particle creation/annihilation operators that define it, one could initially choose $| \Phi_0 \rangle$ and $| \Phi_1
\rangle$ such that $C=D$. In practice, however, this is usually not the case as $| \Phi_0 \rangle$ and $| \Phi_1
\rangle$ are constructed independently from each other and have to be seen as given inputs to a MR calculation.
Finding the transformation that allowed to set $C=D$ would constitute an effort similar to performing the
decomposition (\ref{eq:cano}).

The advantage of the Bloch-Messiah-Zumino decomposition of the transformation $(A,B)$ becomes clear when looking
at the two intermediate sets of quasi-particle operators $\tilde{\alpha}^+_\nu$ and $\tilde{\beta}^+_\nu$
\begin{equation}
\tilde{\alpha}^+_{\nu} \equiv \sum_{\mu} D_{\mu \nu} \alpha^+_\mu \, \, \, , \, \, \, \,
\tilde{\beta}^+_{\nu} \equiv \sum_{\mu} C^{\ast}_{\nu \mu} \, \beta^+_\mu   \,  .
\end{equation}
The two product states $| \Phi_0 \rangle$ and $| \Phi_1 \rangle$ are still vacua for the two sets of
quasi-particle operators $\{\tilde \alpha_{\nu}, \tilde \alpha^+_{\nu}\}$ and $\{\tilde \beta_{\nu}, \tilde
\beta^+_{\nu}\}$, respectively. In addition, the transformation expressing the latter in terms of the former
takes the simple BCS-like form of a special Bogoliubov transformation
\begin{equation}
\label{eq:bcslike}
\tilde{\beta}^+_\nu
=   \bar{A}_{\nu \nu} \, \tilde{\alpha}^+_\nu
  + \bar{B}_{\bar{\nu} \nu} \, \tilde{\alpha}_{\bar{\nu}}
\, .
\end{equation}
The matrices $\bar{A}$ and $\bar{B}$ are $2 \times 2$ block-diagonal (after a suitable rearrangement of the
sequence of indices), with the non-vanishing blocks $\bar{A} (p)$ and  $\bar{B} (p)$ being defined as
\begin{equation}
\bar{A} (p) \equiv \left( \begin{array}{cc}
         \bar{A}_{p p } & 0                       \\
         0              & \bar{A}_{\bar{p} \bar{p}}
         \end{array}
  \right) \, , \qquad
\bar{B} (p) \equiv \left( \begin{array}{cc}
         0                   & \bar{B}_{p \bar{p}} \\
         \bar{B}_{\bar{p} p} & 0
         \end{array}
  \right)  \, ,
\end{equation}
where $\bar{A}_{p p} = \bar{A}_{\bar p \bar p}$ and $\bar{B}_{p \bar p} = - \bar{B}_{\bar p p}$. The
block-diagonal structure of $\bar{A}$ and $\bar{B}$ shows that the quasi-particles $\{\tilde{\alpha}_{\nu},
\tilde{\alpha}^+_{\nu}\}$ come in conjugated pairs $(\nu,\bar{\nu})$, whose nature will be discussed below. As
usual, the block structure can be used to formally separate the corresponding quasi-particle basis into two
halves, even though it is not necessary at this point to specify which quantum number is used to distinguish them.
For the rest of the paper, we introduce the convention that the label $p$ refers to $\nu>0$ while $\bar{p}$ refers
to $\nu<0$. Thus, when $\nu = p (\bar p)$, then $\bar \nu = \bar p (p)$.

\item Standard BCS techniques allow us to express $| \Phi_1 \rangle$ in terms of $| \Phi_0 \rangle$ through
\begin{eqnarray}
\label{eq:BCSqp}
| \Phi_1 \rangle
= \tilde{\mathcal{C}}_{01} \,
  \prod_{p > 0}
  \left(   \bar A^*_{pp}
         + \bar B^*_{p\bar p} \,\tilde{\alpha}^+_p \, \tilde{\alpha}^+_{\bar p}
  \right)
  | \Phi_0 \rangle \, ,
\end{eqnarray}
where $\tilde{\mathcal{C}}_{01}$ is a complex normalization factor whose properties will be discussed in
Sec.~\ref{sect:overlap:contra}.
\end{enumerate}
There are a few further remarks to be made before we continue. When $\bar{A}^{\ast}_{pp} \neq 0$ for all $p$,
Eq.~(\ref{eq:BCSqp}) is nothing but Eq.~(E.39) of Ref.~\cite{ring80a}. However, the form above is also valid when $|
\Phi_0 \rangle$ and $| \Phi_1 \rangle$ are orthogonal, i.e. when at least one pair $(p,\bar{p})$ is such that $\bar
A^*_{pp}=0$ (i.e.\ $\bar B^*_{p\bar p}=1$). This situation corresponds to Eq.~(E.38) of Ref.~\cite{ring80a} with $n$
even and "blocked" quasi-particles coming in pairs\footnote{The conjugated
pairs  $\{\tilde{\alpha}^{\dagger}_{\nu}, \tilde{\alpha}^+_{\bar{\nu}}\}$ are \emph{not} necessarily related
through time-reversal transformation.} $(p, \bar p)$. One such case occurs when $| \Phi_1 \rangle$ is obtained as
a two-quasiparticle excitation $(p_{0}, \bar p_{0})$ on top of $| \Phi_0 \rangle$. This amounts to having
$(\bar{A}_{pp} = 1, \bar{B}_{p \bar p}=0)$ for all pairs but the pair $(p_{0}, \bar p_{0})$ for which
 $(\bar{A}_{p_{0}p_{0}} = 0, \bar{B}_{p_{0} \bar p_{0}}=1)$. In
fact, there are two different situations to consider depending on the properties of the matrix $D$. If the latter
is the unit matrix, the two-quasiparticle states created are conjugated \emph{in the original basis}
$\{\alpha_{\mu}, \alpha^{+}_{\mu}\}$ used to characterize $| \Phi_0 \rangle$. However, other two-quasi-particle
excitations on top of $| \Phi_0 \rangle$ can be generated by considering a non-trivial $D$ transformation.

The above procedure to express $| \Phi_1 \rangle$ with respect to $| \Phi_0 \rangle$ is close to the method
discussed in Ref.~\cite{nee83}, although the latter relies on the Thouless representation of quasi-particle
vacua~\cite{thouless61a} which requires to take a more explicit care of orthogonal states.
Refs.~\cite{nee83,bur95,Dob00} are also dedicated to finding a canonical representation appropriate to
non-diagonal overlaps. However, they focus on the more difficult task of finding a basis where both vacua are
\emph{simultaneously} put into a canonical form. In the present work, we are only interested in putting the
Bogoliubov transformation $(A,B)$ describing the transition between the two vacua into a canonical form but not
the vacua themselves.

%
%
\subsection{Overlaps and one/two-body contractions}
\label{sect:overlap:contra}

The form given by Eq.~(\ref{eq:BCSqp}) is very convenient to express the overlap and the matrix element of operators
between the states $| \Phi_0 \rangle$ and $| \Phi_1 \rangle$. First, let us precise the nature of the
normalization constant $\tilde{\mathcal{C}}_{01}$. As already said, the two vacua are usually taken to be normalized
through an appropriate choice of $|\mathcal{C}_{0}|$ and $|\mathcal{C}_{1}|$ in Eq.~(\ref{initialstates}). Calculating
\begin{equation}
\langle \Phi_1  | \Phi_1 \rangle
= | \tilde{\mathcal{C}}_{01} |^2 \, \langle \Phi_0 | \Phi_0 \rangle
  \prod_{p>0} \Big( |\bar{A}_{pp}|^2 + | \bar{B}_{p\bar p}  |^2 \Big)
\end{equation}
from Eq.~(\ref{eq:BCSqp}) and using that $| \bar{A}_{pp}|^2 + | \bar{B}_{p\bar p}  |^2 =1$ for all $p$, the
normalization of the two vacua leads to $| \tilde{\mathcal{C}}_{01}  |^2 = 1$. However, the latter result does not fix
the phase of $\tilde{\mathcal{C}}_{01}$ which remains when calculating the norm overlap between the two vacua
\begin{equation}
\label{eq:overc}
\langle \Phi_0 | \Phi_1 \rangle
= \tilde{\mathcal{C}}_{01} \, \prod_{p>0} \bar A^*_{pp}
= \tilde{\mathcal{C}}_{01} \sqrt{\det(\bar{A}^*)} \, .
\end{equation}
Eq.~(\ref{eq:overc}) differs from the well-known Onishi-Yoshida formula~\cite{onishi66} by that phase factor
$\tilde{\mathcal{C}}_{01}$ which depends on the chosen convention, that is, on the explicit form of the states $|
\Phi_0 \rangle$ and $| \Phi_1 \rangle$ taken as an input of the MR calculation. We will show explicitly in
Sec.~\ref{PNR} how this works for PNR where the phase $\tilde{\mathcal{C}}_{01}$ can be obtained analytically
for all pairs of vacua involved in the calculation. Unfortunately, this is not true for more involved MR
calculations, as for example when superposing particle-number and angular-momentum restorations together with
GCM-type configuration mixings. In such a case, one must design numerical methods to follow the phase
$\tilde{\mathcal{C}}_{01}$ \cite{Har82a,nee83,Ena99a,valor00c}.

For reasons that will become clear in a moment, we define for
$p\neq q$ the quantities
\begin{eqnarray}
\label{eq:define_nu}
\langle \Phi_0 | \Phi_1, p \rangle
& \equiv & \tilde{\mathcal{C}}_{01} \,
           \prod_{p' >0 \atop p'\neq p} \, \bar A^*_{p'p'} \,  ,
           \\
\label{eq:define_nu2}
\langle \Phi_0 | \Phi_1, p, q \rangle
& \equiv & \tilde{\mathcal{C}}_{01} \, \prod_{p' >0 \atop p'\neq p,q} \, \bar A^*_{p'p'} \, ,
\end{eqnarray}
using the notations of Ref.~\cite{donau98}. In Eqs.~(\ref{eq:define_nu}-\ref{eq:define_nu2}), $\langle \Phi_0 |
\Phi_1, p \rangle$ and $\langle \Phi_0 | \Phi_1, p, q \rangle$ remain unchanged when substituting $p$ and/or $q$
by $\bar p$ and/or $\bar q$ in such a way that one can refer to $\langle \Phi_0 | \Phi_1, \nu \rangle$ and $\langle \Phi_0 | \Phi_1, \nu, \mu \rangle$. Note that the latter overlap does not need to be defined for $\nu=\mu$ or $\nu=\bar{\mu}$ as it cannot appear in such cases. Indeed, this would correspond to removing a conjugated pair twice from $| \Phi_1 \rangle$, which would lead to

\begin{eqnarray}
| \Phi_1 , \nu , \nu \rangle &\propto& \tilde \alpha_\nu \tilde \alpha_{\bar \nu} | \Phi_1 , \nu \rangle = 0 \, \, \, .
\end{eqnarray}
If needed, such particular cases can be trivially taken into account by extending the notation introduced in Eq.~(\ref{eq:define_nu2}) through
\begin{equation}
\label{eq:phi0:phi1:nu:nu}
\langle \Phi_0  | \Phi_1, \nu, \nu \rangle = \langle \Phi_0 | \Phi_1, \nu, \bar \nu \rangle
\equiv 0 \, \, .
\end{equation}
Starting from Eq.~(\ref{eq:BCSqp}), it is now straightforward to calculate one-body contractions in the
quasi-particle basis $\{\tilde \alpha_{\mu}, \tilde \alpha^+_\mu\}$ using the SWT. Using the notations introduced in Eqs.~\ref{eq:overc} and~\ref{eq:define_nu}, one finds that
\begin{eqnarray}
\label{eq:expone}
\langle \Phi_0 | \tilde \alpha^+_\nu \tilde \alpha_{\mu}
| \Phi_1 \rangle
& = & \langle \Phi_0 | \tilde \alpha^+_\nu \tilde \alpha^+_{\mu} | \Phi_1 \rangle
  =   0
      \, ,
      \nonumber \\
\langle \Phi_0 | \tilde \alpha_\nu \tilde \alpha^+_{\mu} | \Phi_1 \rangle  \, & = & \delta_{\nu \mu} \langle
\Phi_0 | \Phi_1 \rangle  \, ,
      \nonumber \\
\langle \Phi_0 | \tilde \alpha_\nu \tilde \alpha_\mu | \Phi_1 \rangle  \,
& = & \delta_{\bar \nu \mu }  \, \bar B^*_{\bar \nu \nu}  \,
      \langle \Phi_0 | \Phi_1, \nu \rangle
\, .
\end{eqnarray}
All two-body contractions are obtained in the same manner and can be written in a compact way using the notation introduced in Eqs.~\ref{eq:define_nu2} and~\ref{eq:phi0:phi1:nu:nu}, e.g.
\begin{eqnarray}
\label{eq:exptwo}
\lefteqn{
\langle \Phi_0 | \tilde \alpha_\nu \tilde  \alpha_\mu \tilde
\alpha_\gamma \tilde  \alpha_\delta | \Phi_1 \rangle
} \nonumber \\
&=& \delta_{\bar \nu \mu }  \, \delta_{\bar \gamma \delta } \,
    \bar B^*_{\bar \nu \nu }  \, \bar B^*_{\bar \gamma \gamma } \,
    \langle \Phi_0 | \Phi_1,\nu,\gamma  \rangle
    \nonumber \\
& &-\delta_{\bar \nu \gamma  }  \, \delta_{\bar \mu \delta } \,
    \bar B^*_{\bar \nu  \nu}  \, \bar B^*_{\bar \mu \mu} \,
    \langle \Phi_0 | \Phi_1,\nu,\mu \rangle
    \nonumber \\
& &+\delta_{\bar \nu \delta}  \, \delta_{\bar \mu \gamma} \,
    \bar B^*_{\bar \nu \nu}  \, \bar B^*_{\bar \mu \mu}  \,
    \langle \Phi_0 | \Phi_1,\nu,\mu\rangle \, .
\end{eqnarray}
%
%
\subsection{More convenient bases}

In order to take advantage of the results obtained in the previous
section, one has to express the single-particle creation and
annihilation operators $\{a_{i}, a^+_i\}$ in terms of the quasi-particle
operators $\{\tilde\alpha_\nu, \tilde \alpha^+_\nu\}$. Introducing
the matrices $\tilde{U}^0 = U^{0} D$ and $\tilde{V}^0 = V^{0} D$, we
have
\begin{equation}
\label{eq:ai_atilde}
a^+_i
= \sum_{\nu}
  \big(   \tilde{U}^{0*}_{i\nu} \, \tilde{\alpha}^+_\nu
        + \tilde{V}^0_{i\nu}    \, \tilde{\alpha}_\nu
  \big) \, .
\end{equation}
Many formulas derived below will simplify by introducing the quasi-particle wave-function associated with the
quasi-particle operators $\{\tilde \alpha_\nu,  \tilde \alpha^+_\nu\}$
\begin{equation}
\left(
\begin{array} {c}
\tilde \alpha^+_\nu \\
\tilde \alpha_\nu
\end{array}
\right) \, | 0 \rangle \equiv \left(
\begin{array} {c}
| \phi_{\nu} \rangle
   \\
| \varphi_{\bar \nu} \rangle\end{array} \right) \, ,
\end{equation}
where the choice of labeling the lower component $| \varphi_{\bar \nu} \rangle$ with the quantum number of the
conjugated state in the other half of the basis is dictated by convenience, as will become clearer in the PNR case
discussed in Sec.~\ref{PNR}.

The upper and lower components can be expressed in terms of the arbitrary single-particle basis $\{ a_i, a^+_i\}$
through
\begin{eqnarray}
| \phi_\nu \rangle
& = & \sum_{i} \tilde U^{0 \, T}_{\nu i} \, | i \rangle \, , \label{eq:def_sp2}
      \\
| \varphi_{\bar \nu} \rangle
& = & \sum_{i} \tilde V^{0 \, +}_{\nu i} \, | i \rangle \, . \label{eq:def_sp1}
\end{eqnarray}
We recall that $\tilde{U}^0$ and $\tilde{V}^{0}$ introduced above involve a unitary transformation $D$ determined
by the Bloch-Messiah-Zumino decomposition of the Bogoliubov transformation $(A,B)$. These definitions will allow
us to use matrix elements of the effective vertices $\bar{v}^{\rho\rho}$ and $\bar{v}^{\kappa\kappa}$ written in
the mixed basis of upper and lower components of the quasi-particle states $\{| \phi_\nu \rangle, | \varphi_{\bar
\nu} \rangle\}$, e.g.
\begin{eqnarray}
\label{newmatrixelements}
\bar{v}^{\rho\rho}_{\varphi_{\nu}\varphi_{\mu} \phi_{\lambda }\phi_{\gamma}}
&\equiv & \langle \varphi_{\nu}\varphi_{\mu} | \bar{v}^{\rho\rho} | \phi_{\lambda}\phi_{\gamma} \rangle
          \nonumber \\
& = & \sum_{ijkl} \tilde V^0_{i\bar \nu} \, \tilde V^0_{j\bar \mu} \,
      \tilde U^0_{k\lambda} \, \tilde U^0_{l\gamma} \, \bar {v}^{\rho\rho}_{ijkl} \, ,
\end{eqnarray}
which are at variance with those expressed in terms of roman indices $(i,j,k,l)$ which relate to the initial
arbitrary basis $\{a_{i}, a^{+}_{i}\}$.

\subsection{Energy kernel from the SWT}
\label{sec:enerkernelSWT1}

Thanks to Eqs.~(\ref{eq:expone}) and~(\ref{eq:exptwo}), the contribution $\langle \Phi_0| v_{12} | \Phi_1
\rangle/\langle \Phi_0 | \Phi_1 \rangle$ to the energy kernel $E_{SWT}[0,1]$ calculated through the application of the SWT reads as
\begin{widetext}
\begin{alignat}{3}
\label{eneroverlap}
{E}^{\rho\rho}_{SWT}[0,1] + {E}^{\kappa\kappa}_{SWT}[0,1]
 =
&  \frac{1}{2} \sum_{\nu \mu} \bar{v}^{}_{\varphi_\nu \varphi_\mu \varphi_\nu \varphi_{\mu}} & &  + \frac{1}{4}
\sum_{\nu \mu} \bar{v}^{}_{\varphi_\nu \phi_{\bar \nu} \varphi_\mu \phi_{\bar \mu}}
      \nonumber  \\
+ &  \frac{1}{2} \sum_{\nu \mu} \bar{v}^{}_{\varphi_\nu \varphi_\mu \phi_{\nu} \varphi_{\mu}} \,
     \bar B^*_{\nu \bar \nu }
      \frac{\langle \Phi_0 | \Phi_1,\nu \rangle}{\langle \Phi_0 |\Phi_1 \rangle}  &
&  + \frac{1}{4} \sum_{\nu \mu} \bar{v}^{}_{\varphi_\nu \varphi_{\bar \nu}\varphi_\mu \phi_{\bar \mu}} \,
      \bar B^*_{\nu \bar \nu }
      \frac{\langle \Phi_0 | \Phi_1,\nu \rangle}{\langle \Phi_0 | \Phi_1 \rangle}
     \nonumber \\
+ &  \frac{1}{2} \sum_{\nu \mu} \bar{v}^{}_{\varphi_\mu \varphi_\nu \varphi_\mu \phi_{ \nu}} \,
     \bar B^*_{ \nu \bar \nu}
      \frac{\langle \Phi_0 | \Phi_1,\nu \rangle}{\langle \Phi_0 | \Phi_1 \rangle}  &
&  + \frac{1}{4} \sum_{\nu \mu} \bar{v}^{}_{\varphi_\mu \phi_{\bar\mu} \phi_{ \nu} \phi_{\bar \nu}} \,
      \bar B^*_{ \nu \bar \nu}
      \frac{\langle \Phi_0 | \Phi_1,\nu \rangle}{\langle \Phi_0 | \Phi_1 \rangle}
      \nonumber  \\
+ &  \frac{1}{2} \sum_{\nu \mu \atop \nu \neq \mu, \bar \mu} \bar{v}^{}_{\varphi_\nu \varphi_\mu \phi_{\nu} \phi_{\mu}} \, \bar B^{*}_{ \nu
\bar \nu} \bar B^{*}_{ \mu \bar \mu}
      \frac{\langle \Phi_0 | \Phi_1,\nu,\mu \rangle}{\langle \Phi_0 | \Phi_1 \rangle}
 &
&  + \frac{1}{4} \sum_{\nu \mu \atop \nu \neq \mu, \bar \mu} \bar{v}^{}_{\varphi_\nu \varphi_{\bar \nu} \phi_{ \mu} \phi_{\bar \mu}} \,
 \bar B^*_{\nu \bar \nu } \bar B^{*}_{\mu \bar \mu }
      \frac{\langle \Phi_0 | \Phi_1,\nu,\mu \rangle}{\langle \Phi_0 | \Phi_1 \rangle}
\, .
\end{alignat}
\end{widetext}
The terms in Eq.~(\ref{eneroverlap}) are ordered in such a manner that those in the left column can be later
identified with terms bilinear in $\rho^{01}$ when using the GWT, whereas the terms in the right column are
related to terms proportional to $\kappa^{10*} \kappa^{01}$. It is essential for the following to note that
terms with $\nu=\mu$ or $\nu = \bar \mu$ do not appear in the last line of Eq.~(\ref{eneroverlap}), i.e. they are zero.

%
\subsection{Energy kernel from the GWT}
\label{sec:transdens}

The use of the SWT for the calculation of MR energy kernels relies on the construction of the very particular
basis $\{\tilde \alpha_\nu, \tilde \alpha^+_\nu\}$. All practical applications, however, have used the
GWT~\cite{balian69a} which typically provides, in any arbitrary single-particle basis, energy kernels of the form
given by Eq.~(\ref{eq:hamil_phi01}).

The next step is to compare the expressions of the energy kernel obtained from the SWT and the GWT. Using
Eqs.~(\ref{eq:expone}) and~(\ref{eq:ai_atilde}), the transition density matrices are given by
\begin{eqnarray}
\label{lien1}
\rho^{01}_{ji}
& = &  \sum_{\nu} \tilde V^{0*}_{j \nu} \, \tilde V^{0}_{i\nu}
      +\sum_{\nu } \tilde U^{0}_{j \bar \nu}\,  \tilde V^{0}_{i\nu} \, \bar B^{*}_{\bar \nu \nu}
       \frac{\langle \Phi_0 | \Phi_1, \nu \rangle}{\langle \Phi_0 | \Phi_1 \rangle}
     \\
\label{lien2}
\kappa^{10}_{ji}
& = &   \sum_{\nu} \tilde V^{0*}_{j \nu} \, \tilde U^{0}_{i\nu}
      + \sum_{\nu } \tilde U^{0}_{j \bar \nu} \, \tilde U^{0}_{i\nu} \, \bar B^*_{\bar \nu \nu}
        \frac{\langle \Phi_0 | \Phi_1, \nu \rangle}{\langle \Phi_0 | \Phi_1 \rangle}
     \\
\label{lien3}
\kappa^{01 \, \ast}_{ij}
& = &  \sum_{\nu} \tilde V^{0}_{i \nu} \, \tilde U^{0*}_{j\nu}
     + \sum_{\nu } \tilde V^{0}_{i\nu}  \, \tilde V^{0}_{j \bar \nu}  \,
       \bar B^{*}_{ \bar \nu \nu} \frac{\langle \Phi_0  | \Phi_1, \nu \rangle}{\langle \Phi_0 | \Phi_1 \rangle}
\end{eqnarray}
where the running index in the sums refers to the basis $\{\tilde \alpha_\nu, \tilde \alpha^+_\nu\}$. Inserting
Eqs.~(\ref{lien1}-\ref{lien3}) into Eq.~(\ref{eq:hamil_phi01}), the two-body part of the energy kernel is obtained
from the GWT as
\begin{widetext}
\begin{alignat}{4}
\label{eneroverlapdirectGWT}
{E}^{\rho\rho}_{GWT}[0,1] + {E}^{\kappa\kappa}_{GWT}[0,1]
 =
 & \frac{1}{2} \sum_{\nu \mu} \bar{v}^{}_{\varphi_\nu \varphi_\mu \varphi_\nu \varphi_{\mu}} &
&  + \frac{1}{4} \sum_{\nu \mu} \bar{v}^{}_{\varphi_\nu \phi_{\bar \nu} \varphi_\mu \phi_{\bar \mu}}
      \nonumber  \\
+ &  \frac{1}{2} \sum_{\nu \mu} \bar{v}^{}_{\varphi_\nu \varphi_\mu \phi_{\nu} \varphi_{\mu}} \,
     \bar B^*_{\nu \bar \nu }
      \frac{\langle \Phi_0 | \Phi_1,\nu \rangle}{\langle \Phi_0 |\Phi_1 \rangle}  &
&  + \frac{1}{4} \sum_{\nu \mu} \bar{v}^{}_{\varphi_\nu \varphi_{\bar \nu}\varphi_\mu \phi_{\bar \mu}} \,
      \bar B^*_{\nu \bar \nu }
      \frac{\langle \Phi_0 | \Phi_1,\nu \rangle}{\langle \Phi_0 | \Phi_1 \rangle}
     \nonumber \\
+ &  \frac{1}{2} \sum_{\nu \mu} \bar{v}^{}_{\varphi_\mu \varphi_\nu \varphi_\mu \phi_{ \nu}} \,
\bar B^*_{ \nu \bar \nu}
      \frac{\langle \Phi_0 | \Phi_1,\nu \rangle}{\langle \Phi_0 | \Phi_1 \rangle}  &
&  + \frac{1}{4} \sum_{\nu \mu} \bar{v}^{}_{\varphi_\mu \phi_{\bar\mu} \phi_{ \nu} \phi_{\bar \nu}} \,
      \bar B^*_{ \nu \bar \nu}
      \frac{\langle \Phi_0 | \Phi_1,\nu \rangle}{\langle \Phi_0 | \Phi_1 \rangle}
      \nonumber  \\
+ &  \frac{1}{2} \sum_{\nu \mu} \bar{v}^{}_{\varphi_\nu \varphi_\mu \phi_{\nu} \phi_{\mu}} \, \bar B^{*}_{ \nu
\bar \nu} \bar B^{*}_{ \mu \bar \mu} \frac{\langle \Phi_0 | \Phi_1, \nu \rangle}
     {\langle \Phi_0 | \Phi_1 \rangle}
\frac{\langle \Phi_0 | \Phi_1, \mu \rangle}
     {\langle \Phi_0 | \Phi_1 \rangle}
 &
&  + \frac{1}{4} \sum_{\nu \mu} \bar{v}^{}_{\varphi_\nu \varphi_{\bar \nu} \phi_{ \mu} \phi_{\bar \mu}} \,
 \bar B^*_{\nu \bar \nu } \bar B^{*}_{\mu \bar \mu }
\frac{\langle \Phi_0 | \Phi_1, \nu \rangle}
     {\langle \Phi_0 | \Phi_1 \rangle}
\frac{\langle \Phi_0 | \Phi_1, \mu \rangle}
     {\langle \Phi_0 | \Phi_1 \rangle}
\, ,
\end{alignat}
\end{widetext}
where each line of Eq.~(\ref{eneroverlapdirectGWT}) can be put in correspondence with those of
Eq.~(\ref{eneroverlap}). Expression (\ref{eneroverlap}) is
recovered term by term from Eq.~(\ref{eneroverlapdirectGWT}) as one can easily check that
\begin{eqnarray}
\label{eq:vvv}
\frac{\langle \Phi_0 | \Phi_1, \nu \rangle}
     {\langle \Phi_0 | \Phi_1 \rangle}
\frac{\langle \Phi_0 | \Phi_1, \mu \rangle}
     {\langle \Phi_0 | \Phi_1 \rangle}
=
\frac{\langle \Phi_0 | \Phi_1, \nu, \mu \rangle}
     {\langle \Phi_0 | \Phi_1 \rangle}
\, ,
\end{eqnarray}
for all pairs $(\mu,\nu)$ such that $\nu\neq\mu$ or
$\nu\neq\bar\mu$. However, one is left with additional terms in the last line of Eq.~(\ref{eneroverlapdirectGWT}), i.e. for $\nu=\mu$ and
$\nu=\bar\mu$, which have no correspondent in the last line of Eq.~(\ref{eneroverlap}).

In spite of such an apparent difference, the energy kernel $E_{GWT}[0,1]$ is strictly equal to $E_{SWT}[0,1]$ when calculated from a genuine Hamilton operator, that is, within the strict projected-GCM
approximation. To prove it, let us restart from last line of Eq.~(\ref{eneroverlapdirectGWT}) and isolate the possible
source of discrepancies (i.e.\ terms with $\nu=\mu$ and/or $\nu = \bar \mu$). Considering first the terms coming
from $\kappa^{10 \, *} \kappa^{01}$, using the antisymmetry of the interaction and the properties of the matrix
$\bar B$, we find that
\begin{eqnarray}
& &
\frac{1}{4} \sum_{\nu} \bar{v}^{}_{\varphi_\nu \varphi_{\bar \nu} \phi_{ \nu} \phi_{\bar \nu}} \,
\bar B^*_{\nu \bar \nu } \bar B^{*}_{\nu \bar \nu }
\frac{\langle \Phi_0 | \Phi_1, \nu \rangle}
     {\langle \Phi_0 | \Phi_1 \rangle}
\frac{\langle \Phi_0 | \Phi_1, \nu \rangle}
     {\langle \Phi_0 | \Phi_1 \rangle}  
     \nonumber \\
& &
     + \frac{1}{4} \sum_{\nu} \bar{v}^{}_{\varphi_\nu \varphi_{\bar \nu} \phi_{\bar  \nu} \phi_{\nu}} \,
 \bar B^*_{\nu \bar \nu } \bar B^{*}_{\bar \nu \nu }
\frac{\langle \Phi_0 | \Phi_1, \nu \rangle}
     {\langle \Phi_0 | \Phi_1 \rangle}
\frac{\langle \Phi_0 | \Phi_1, \nu \rangle}
     {\langle \Phi_0 | \Phi_1 \rangle}
\nonumber \\
&   & \quad =
      \frac{1}{2} \sum_{\nu} \bar{v}^{}_{\varphi_\nu \varphi_{\bar \nu} \phi_{ \nu} \phi_{\bar \nu}} \,
 \bar B^*_{\nu \bar \nu } \bar B^{*}_{\nu \bar \nu }
\frac{\langle \Phi_0 | \Phi_1, \nu \rangle}
     {\langle \Phi_0 | \Phi_1 \rangle}
\frac{\langle \Phi_0 | \Phi_1, \nu \rangle}
     {\langle \Phi_0 | \Phi_1 \rangle} \, ,
\nonumber \\
&   &
\end{eqnarray}
which happens to cancel out exactly the contributions with $\mu = \bar{\nu}$ originating from $\rho^{01} \, \rho^{01}$.
For this cancelation to occur it is crucial that the \emph{same} interaction is employed in both terms, as it is
the case if a genuine Hamiltonian is used. Altogether only the term with $\nu=\mu$ from the $\rho^{01} \,
\rho^{01}$ contributions could not be combined with terms from $\kappa^{10 \, \ast} \, \kappa^{01}$. However,
thanks to the antisymmetry of the two-body interaction, these terms also cancel out.

Therefore, the prerequisites for a complete matching between estimates of the two-body energy kernel
from the SWT and the GWT, independently of the appearance of divergences or not, are that
\begin{itemize}
\item[1.] the same interaction kernel is used in both bilinear terms of the EDF in order to properly recombine specific terms coming
from $\rho^{01} \, \rho^{01}$ and $\kappa^{10 \, \ast} \, \kappa^{01}$.
\item[2.] the antisymmetry of the interaction kernel is properly accounted for.
\end{itemize}
The key point is that none of the two previous conditions are fulfilled in general when constructing a MR-EDF
through an \emph{ansatz} of the form of Eq.~(\ref{eq:e01}): the matrix elements $\bar{v}^{\rho\rho}_{ijkl}$ are not
necessarily antisymmetric, and $\bar{v}^{\rho\rho}_{ijkl}$ and $\bar{v}^{\kappa\kappa}_{ijkl}$ are in general not
the same. Consequently, the kernels $\mathcal{E}_{SWT}[0,1]$ and $\mathcal{E}_{GWT}[0,1]$ defined in the EDF method by analogy with the Hamiltonian formalism are different, with $\mathcal{E}_{GWT}[0,1]$ containing terms that do not appear in $\mathcal{E}_{SWT}[0,1]$.

%
\section{Corrected energy kernel}
\label{correctingMREDF}

The observations of the previous section are at the heart of the problems encountered in MR-EDF calculations that
rely on the GWT to construct the MR energy kernels from the SR-EDF. Let us analyze the situation further and
demonstrate that there exist in fact several levels of problems of different physical origin. This will lead us to
advocate the explicit removal of specific contributions to $\mathcal{E}_{GWT}[0,1]$ in order to perform meaningful
MR-EDF calculations.

It is crucial to note already here that, as the correction procedure proposed in the following is based on an
analogy with the Hamiltonian formalism, it can only be implemented for functionals containing integer powers of
the density matrices. We will come back to that crucial point when discussing the long term implications of the
present work.

%
%
\subsection{Bilinear functional}

Let us start with the bilinear energy kernel $\mathcal{E}_{GWT}[0,1]$ given by Eq.~(\ref{eq:e01}). As already
discussed, generalizing the functional to the multi-reference case using the GWT as a motivation leads to the same structure as
the SR-EDF given in Eq.~(\ref{eq:e00}), except that intrinsic densities have been replaced with transition ones.
Again, the terms $\mathcal{E}^{\rho\rho}_{GWT}[0,1]$ and $\mathcal{E}^{\kappa\kappa}_{GWT}[0,1]$ might contain different
vertices.

We now introduce more compact expressions of the transition density matrices using the identity $\langle \Phi_0  |
\Phi_1 \rangle =\bar A^*_{\nu\nu}\langle \Phi_0  | \Phi_1, \nu \rangle$
\begin{eqnarray}
\label{eq:rho01rho0}
\rho^{01}
& = & \rho^{00} + \tilde{U}^{0}  \, \bar{Z} \, {\tilde{V}^0}{}^T
      \nonumber \\
\kappa^{01}
& = & \kappa^{00} + \tilde{U}^{0} \, \bar{Z} \, {\tilde{U}^0}{}^T
      \nonumber \\
\kappa^{10^*}
& = & \kappa^{00^*} - \tilde{V}^{0}  \, \bar{Z} \, {\tilde{V}^0}{}^T \, ,
\end{eqnarray}
where $\rho^{00}$ and $\kappa^{00}$ are the intrinsic density matrices associated with $| \Phi_0 \rangle$, whereas
the matrix $\bar Z$ has non-vanishing matrix elements of the form $\bar Z_{\bar \nu \nu} =(\bar B_{\bar \nu \nu}
\bar A^{-1}_{\nu \nu})^*$. Note that $Z=D\bar Z D^*=(BA^{-1})^*$ is nothing but the Thouless
matrix~\cite{thouless61a,balian69a} associated with the Bogoliubov transformation connecting  $| \Phi_0 \rangle$ and $| \Phi_1 \rangle$. Equation~(\ref{eq:rho01rho0}) is an alternative form to the one given in
Appendix~E of Ref.~\cite{ring80a}.

Using the mixed basis introduced in Eqs.~(\ref{eq:def_sp2}-\ref{eq:def_sp1}), as well as
Eqs.~(\ref{eneroverlapdirectGWT}) and~(\ref{eq:rho01rho0}), the interaction part of the energy kernel
 can be written as
\begin{widetext}
\begin{alignat}{3}
\mathcal{E}^{\rho\rho}_{GWT}[0,1] + \mathcal{E}^{\kappa\kappa}_{GWT}[0,1]
=
& %
  \frac{1}{2} \sum_{\nu \mu} \bar{v}^{\rho\rho}_{\varphi_\nu \varphi_\mu \varphi_\nu \varphi_{\mu}} &
&  + \frac{1}{4} \sum_{\nu \mu} \bar{v}^{\kappa\kappa}_{\varphi_\nu \phi_{\bar \nu} \varphi_\mu \phi_{\bar \mu}}
\label{kernel1}
 \\
+ &  \frac{1}{2} \sum_{\nu \mu} \bar{v}^{\rho\rho}_{\varphi_\nu \varphi_\mu \phi_{\nu} \varphi_{\mu}} \,
     \bar Z_{\nu \bar\nu } &
&  + \frac{1}{4} \sum_{\nu \mu} \bar{v}^{\kappa\kappa}_{\varphi_\nu \varphi_{\bar \nu}\varphi_\mu \phi_{\bar \mu}}
\,
     \bar Z_{\nu \bar \nu}
\label{kernel2}
\\
+ &  \frac{1}{2} \sum_{\nu \mu} \bar{v}^{\rho\rho}_{\varphi_\mu \varphi_\nu \varphi_\mu \phi_{ \nu}} \,
     \bar Z_{\nu \bar\nu } &
&  + \frac{1}{4} \sum_{\nu \mu} \bar{v}^{\kappa\kappa}_{\varphi_\mu \phi_{\bar\mu} \phi_{ \nu} \phi_{\bar \nu}} \,
      \bar Z_{\nu \bar \nu}
\label{kernel3}
 \\
+ &  \frac{1}{2} \sum_{\nu \mu} \bar{v}^{\rho\rho}_{\varphi_\nu \varphi_\mu \phi_{\nu} \phi_{\mu}} \,
     \bar Z_{ \nu\bar\nu} \, \bar Z_{ \mu \bar\mu} &
&  + \frac{1}{4} \sum_{\nu \mu} \bar{v}^{\kappa\kappa}_{\varphi_\nu \varphi_{\bar \nu} \phi_{ \mu} \phi_{\bar \mu}} \,
     \bar Z_{ \nu \bar\nu } \, \bar Z_{ \mu \bar\mu }
\, , \label{kernel4}
\end{alignat}
\end{widetext}
where the sums run over all states. It is crucial to realize that the 8 terms in Eq.~(\ref{kernel1}-\ref{kernel4})
are arranged in exactly the same manner as the 8 terms in Eq.~(\ref{eneroverlap}).

%
%
\subsubsection{Correcting for divergences and finite steps}
\label{correctGWT}

By analogy with the strict projected-GCM case discussed in Sec.~\ref{illdefined}, the contribution to
$\mathcal{E}_{GWT}[0,1]$ from the terms $\nu=\mu$ and $\nu=\bar \mu$ in the last line of Eq.~(\ref{kernel4}) should be zero.
However, the contribution encoded into the bilinear term $\rho^{01} \rho^{01}$ is not zero for $\mu=\nu$ in
line~(\ref{kernel4}) when the effective vertex $\bar{v}^{\rho\rho}$ is not antisymmetric. Also, the vertices
$\bar{v}^{\rho\rho}$ and $\bar{v}^{\kappa\kappa}$ have no reason to be the same in an EDF approach; as a result,
the contributions from $\rho^{01} \rho^{01}$ for $\mu=\bar{\nu}$ and $\kappa^{10 \, \ast} \kappa^{01}$ for ($\mu = \nu$, $\mu = \bar
\nu$) do not cancel out anymore in line~(\ref{kernel4}).

This difficulty is caused by the use of the GWT as a motivation to construct MR energy kernel from the SR EDF.
However, it must be made clear that the GWT itself cannot and should not be blamed for this failure, as it is
stretched beyond its applicability when used to motivate the form of an energy kernel that does not correspond to
the matrix element of an operator. It is problematic though as the terms that do not cancel out in
line~(\ref{kernel4}) are at the origins of the divergencies and steps seen in MR-EDF calculations based on the
GWT~\cite{doba05a} as discussed in detail in Ref.~\cite{bender07x}.

As a matter of illustration, we now focus on divergences. Divergences will occur in $\mathcal{E}_{GWT}[0,1]
\langle \Phi_0 | \Phi_1 \rangle$ if the two vacua involved are orthogonal; i.e.\ when it exists $| \Phi_0 \rangle$
and $| \Phi_1 \rangle$ in the MR set such that $\langle \Phi_0  | \Phi_1 \rangle =0$. As can be seen from
Eq.~(\ref{eq:overc}), the overlap $\langle \Phi_0 | \Phi_1 \rangle$ is zero if at least one of the matrix elements $
\bar A^*_{pp}$ is zero. Because the non-zero contribution to $\mathcal{E}_{GWT}[0,1]
\langle \Phi_0 | \Phi_1 \rangle$ coming from the term involving the pair $(p,\bar{p})$ in Eq.~(\ref{kernel4}) is proportional to
\begin{equation}
\label{diverge}
\left(\bar Z_{ p \bar p }\right)^{2} \, \langle \Phi_0  | \Phi_1 \rangle
= \tilde{\mathcal{C}}_{01} \, \left(\bar B^{\ast}_{p\bar p}\right)^{2} \,
  \frac{1}{\bar A^*_{pp}} \, \prod_{p'>0 \atop p'\neq p} \, \bar A^*_{p'p'}
\, ,
\end{equation}
it will diverge as $\bar A^*_{pp}$ goes to zero, except if another factor $\bar{A}^*_{p'p'}$ in the numerator
happens to be zero as well. For particular MR calculations and very specific situations, it is indeed possible that
several conjugated pairs $(p,\bar{p}), (p',\bar{p}'), \ldots$ are such that $\bar A^*_{pp} = \bar A^*_{p'p'} =
\ldots =0$ at the same time. In such a particular case, there will be no divergence of the MR-energy due to dangerous terms
in Eq.~(\ref{kernel4}).

A regularized energy kernel is obtained by removing the spurious contributions to $\mathcal{E}^{\rho\rho}_{GWT}$ and
$\mathcal{E}^{\kappa\kappa}_{GWT}$ that constitute the difference between $\mathcal{E}_{GWT}[0,1]$ and
$\mathcal{E}_{SWT}[0,1]$. This amounts to removing terms involving only one conjugated pair $(p,\bar p )$ at a time in
Eq.~(\ref{kernel4})
\begin{eqnarray}
\label{eq:cor_rr} \mathcal{E}^{\rho \rho}_{CG}[0,1] & = & \frac{1}{2} \sum_{p>0}
      \Big( \bar{v}^{\rho\rho}_{\varphi_{p}\varphi_{p}\phi_{p}\phi_{p}}
            +\bar{v}^{\rho\rho}_{\varphi_{\bar{p}}\varphi_{\bar{p}}\phi_{\bar{p}}\phi_{\bar{p}}}
      \nonumber \\
&   &
            -\bar{v}^{\rho\rho}_{\varphi_{p}\varphi_{\bar{p}}\phi_{p}\phi_{\bar{p}}}
            -\bar{v}^{\rho\rho}_{\varphi_{\bar{p}}\varphi_{p}\phi_{\bar{p}}\phi_{p}} \Big) \left(\bar Z_{ p \bar p }\right)^{2}
       \, ,
       \\
\label{eq:cor_kk} \mathcal{E}^{\kappa \kappa}_{CG}[0,1] & = & \sum_{p>0}
\bar{v}^{\kappa\kappa}_{\varphi_{p}\varphi_{\bar{p}}\phi_{p}\phi_{\bar{p}}} \,
       \left(\bar Z_{ p \bar p }\right)^{2}
\, ,
\end{eqnarray}
where the sum runs over $p=\nu>0$ only because all the matrix elements involving a given conjugated pair have been
explicitly spelled out.

Some important comments should be made
\begin{itemize}
\item[(i)]
The sum of Eqs.~(\ref{eq:cor_rr}) and~(\ref{eq:cor_kk}) is of course zero if the MR energy
kernel is obtained as the expectation value of the Hamiltonian, while it is non zero for a general functional energy
kernel.
\item[(ii)]
The correction established through Eqs.~(\ref{eq:cor_rr}) and~(\ref{eq:cor_kk}) is zero when $| \Phi_0 \rangle = |
\Phi_1 \rangle$. Indeed, the Bogoliubov transformation connecting the two states is trivial in this case with $A$
being the unit matrix and $B$ being zero. As a result, the correction does not modify the diagonal kernels
$\mathcal{E}[0,0]$, which preserves the continuity between the SR functional and MR energy kernels
$\mathcal{E}[\rho^{00},\kappa^{00},\kappa^{00 \, \ast}] = \mathcal{E}[0,0]$.
\item[(iii)]
When pairing is not considered explicitly in the theory, the only terms to remove from Eq.~(\ref{kernel4}) are those
obtained from $\rho^{01} \rho^{01}$ for $\mu = \nu$. As already discussed, such terms are indeed different from
zero if the vertex $\bar{v}^{\rho\rho}$ is not antisymmetric. As a matter of fact, we believe these terms to be
responsible for the divergences seen recently in the restoration of angular momentum from cranked HF
states~\cite{doba06a}; see Sec.~\ref{sectionzeropairing}.
\item[(iv)]
The correction to $\mathcal{E}_{GWT}[0,1]$ is independent of the normalization factor $\tilde{\mathcal{C}}_{01}$ and
therefore of the phase conventions chosen to define the two vacua $| \Phi_0 \rangle$ and $| \Phi_1 \rangle$. This
is so because the ratio $\bar B^*_{\bar p p}/\bar A^*_{p p}$ is independent of $\tilde{\mathcal{C}}_{01}$. However, it
should be kept in mind that the kernel $\mathcal{E}_{GWT}[0,1]$ is to be multiplied eventually by $\langle \Phi_0  |
\Phi_1 \rangle$ where $\tilde{\mathcal{C}}_{01}$ enters explicitly.
\item[(v)]
Equations.~(\ref{eq:cor_rr}) and~(\ref{eq:cor_kk}) not only remove possible poles leading to divergences but also correct
the energy kernel \emph{away} from those potential poles. It is a crucial result of the present work to realize
that current calculations are not only compromised through divergences but also through finite spurious
contributions.
\item[(vi)]
As just said, poles associated to zeros of the norm overlaps are only a part of the problem. From that point of view, the study of Ref.~\cite{oi05a} about the nodal lines of the
overlap between cranked HFB states rotated by different values of the Euler angles is of prime interest. In
Ref.~\cite{oi05a}, it is shown in particular that the structure of nodal lines becomes richer as the states in the
MR set are cranked to higher spins.
\end{itemize}

\subsubsection{Correcting for self-interaction processes}

In the last section, we have isolated the spurious contributions which are specific to $\mathcal{E}_{GWT}[0,1]$ when
it is not obtained as the matrix element of the Hamiltonian. In particular, it was argued that the construction of
MR energy kernels based on the SWT is safe from divergences and steps. However, there exist others, less
dangerous, spurious contributions which are common to $\mathcal{E}_{SWT}[0,1]$ and $\mathcal{E}_{GWT}[0,1]$. This
underlines the fact that the SWT is also stretched beyond its applicability when used to construct EDFs that do
not correspond to the matrix element of a genuine operator.

The first of those problems relates to the fact that the matrix elements $\bar{v}^{\rho\rho}$ might not be
antisymmetrized. This leads to spurious \emph{self-interaction} processes in the functional; i.e.\ the fact that a
nucleon interacts with itself~\cite{perdew81a}. The spurious terms not already removed by the correction discussed
in the previous section, and which should be zero in $\mathcal{E}[0,1]$ are obtained for $\nu = \mu = p$ or $\nu =
\mu = \bar p$ in the terms that are independent of $\bar Z$ [line (\ref{kernel1})] or linear in $\bar Z$ [second and
third line (\ref{kernel2}-\ref{kernel3})]
\begin{eqnarray}
\label{selfint}
\mathcal{E}^{\rho\rho}_{SI}[0,1]
&\equiv& \frac{1}{2} \sum_{p>0}
    \Big(   \bar{v}^{\rho\rho}_{\varphi_p \varphi_p \varphi_p \varphi_{p}}
          + \bar{v}^{\rho\rho}_{\varphi_{\bar p} \varphi_{\bar p} \varphi_{\bar p} \varphi_{\bar p}}
    \Big)
    \nonumber \\
& + &  \frac{1}{2} \sum_{p>0}
    \Big( \bar{v}^{\rho\rho}_{\varphi_p \varphi_p \phi_{p} \varphi_{p}}
         - \bar{v}^{\rho\rho}_{\varphi_{\bar p} \varphi_{\bar p} \phi_{\bar p} \varphi_{\bar p}}
    \Big) \, \bar Z_{ p \bar p }
       \nonumber \\
& + & \frac{1}{2} \sum_{p>0}
    \Big(   \bar{v}^{\rho\rho}_{\varphi_p \varphi_p \varphi_p \phi_{p}}
          - \bar{v}^{\rho\rho}_{\varphi_{\bar p} \varphi_{\bar p} \varphi_{\bar p} \phi_{\bar p}}
    \Big) \, \bar Z_{ p \bar p }
 \, .
\end{eqnarray}
It is clear that these terms are unphysical, as they violate Pauli's principle. This gives a strong motivation to
remove them from the energy functional, as is sometimes done in DFT for electronic systems
\cite{perdew81a,koch01,Ull00aDFT,Leg02aDFT,Ruz07aDFT}, or to construct energy functionals that are
self-interaction-free in the first place \cite{Tao03aDFT}. When using the standard correction,
this will modify the SR energy density functional to what is called an "orbital-dependent functional"
in electronic DFT~\cite{engel03a,kummel08a}, and will lead to
equations of motion that are more difficult to solve numerically than the usual ones. From a
phenomenological point of view, the merits of self-interaction corrected energy functionals for electronic DFT
have not yet been fully clarified, as they improve some observables, but may degrade others when compared to
uncorrected functionals~\cite{Ruz07aDFT}. It has also been pointed out that correcting for the one-body
self-interaction might not be sufficient, as there also might be also $n$-body self-interactions
as well~\cite{Ruz07aDFT,Vydrov07aDFT}.

In the nuclear context, the existence of spurious self-interactions in commonly used energy density functionals
has been pointed out before~\cite{stringari78a,Rutz99a,bender03b}, but was never studied in any detail in the
published literature.
%
%
\subsubsection{Correcting for self-pairing processes}

The second category of problems common to $\mathcal{E}_{SWT}[0,1]$ and $\mathcal{E}_{GWT}[0,1]$ is less obvious to
isolate. In fact, it differs from the two previous sources of problems in the sense that it is subject to
interpretation.

It is based on the observation that in the strict HFB or projected/GCM methods, where the same vertex enters
$E^{\rho\rho}[0,1]$ and $E^{\kappa\kappa}[0,1]$, the terms corresponding to $(\nu = p \, ; \mu = \bar p)$ and
$(\nu = p \, ; \mu = p)$ recombine in Eqs.~(\ref{kernel1}-\ref{kernel3}). As explained in Ref.~\cite{bender07x},
such a recombination of terms can be interpreted as the fact that two nucleons in a pair of conjugated states $(p,\bar p)$ cannot
gain additional binding energy, as compared to the same EDF calculation with no explicit account of pairing,
by scattering onto itself. Such a spurious \emph{self-pairing} process should in principle be excluded from the
functional, although the actual impact of these terms on observables is not clear yet.

Compared to GWT-related and self-interaction problems, removing the self-pairing energy does not amount to putting
specific terms to zero. On the contrary, the reasoning is to allow certain terms to recombine and to provide a
finite, non-zero contribution to the energy kernel. It is the remaining finite contribution which is subject to
interpretation. Here, we follow the argument from Ref.~\cite{bender07x} mentioned above that takes the calculation
without explicit treatment of pairing as a reference point. As a result, the advocated correction amounts to
replacing specific matrix elements of $\bar v^{\kappa\kappa}$ by the corresponding ones of $\bar
v^{\rho\rho}$. Eventually, the correction for self-pairing amounts to the subtraction of
\begin{widetext}
\begin{eqnarray}
\label{selfpair} \mathcal{E}^{\kappa\kappa}_{SP}[0,1] &\equiv& \frac{1}{4} \sum_{p>0} \Big[ \left(
\bar{v}^{\kappa\kappa}_{\varphi_p \phi_{\bar p} \varphi_p \phi_{\bar p}} + \bar{v}^{\kappa\kappa}_{\varphi_{\bar
p}  \phi_p \varphi_{\bar p} \phi_p} - \bar{v}^{\kappa\kappa}_{\varphi_p \phi_{\bar p} \phi_{p} \varphi_{\bar p}} -
\bar{v}^{\kappa\kappa}_{\varphi_{\bar p} \phi_p \phi_{\bar p} \varphi_p}\right) \nonumber \\
&& \hspace{0.9cm}- \left( \bar{v}^{\rho\rho}_{\varphi_p \phi_{\bar p} \varphi_p \phi_{\bar p}} +
\bar{v}^{\rho\rho}_{\varphi_{\bar p}  \phi_p \varphi_{\bar p} \phi_p} - \bar{v}^{\rho\rho}_{\varphi_p \phi_{\bar
p} \phi_{p} \varphi_{\bar p}} - \bar{v}^{\rho\rho}_{\varphi_{\bar p} \phi_p \phi_{\bar p} \varphi_p}\right)\Big]
 \nonumber \\
&+& \frac{1}{4} \sum_{p>0} \Big[ \left(\bar{v}^{\kappa\kappa}_{\varphi_p \varphi_{\bar p} \varphi_p \phi_{\bar
p}} + \bar{v}^{\kappa\kappa}_{\varphi_{\bar p} \varphi_p \phi_{\bar p} \varphi_p } -
\bar{v}^{\kappa\kappa}_{\varphi_p \varphi_{\bar p} \phi_{ p} \varphi_{\bar p}}
-\bar{v}^{\kappa\kappa}_{\varphi_{\bar p} \varphi_p \varphi_{\bar p}
\phi_p} \right)  \nonumber \\
&&\hspace{0.9cm}- \left(\bar{v}^{\rho\rho}_{\varphi_p \varphi_{\bar p} \varphi_p \phi_{\bar p}} +
\bar{v}^{\rho\rho}_{\varphi_{\bar p} \varphi_p \phi_{\bar p} \varphi_p } - \bar{v}^{\rho\rho}_{\varphi_p
\varphi_{\bar p} \phi_{ p} \varphi_{\bar p}} -\bar{v}^{\rho\rho}_{\varphi_{\bar p} \varphi_p \varphi_{\bar p}
\phi_p}
\right) \Big] \, \bar Z_{p \bar p} \nonumber \\
&+& \frac{1}{4} \sum_{p>0} \Big[ \left( \bar{v}^{\kappa\kappa}_{\varphi_p \phi_{\bar p} \phi_{p} \phi_{\bar p}} -
\bar{v}^{\kappa\kappa}_{\varphi_{\bar p} \phi_{p} \phi_{\bar p} \phi_{p}}  + \bar{v}^{\kappa\kappa}_{\phi_{\bar p}
\varphi_p \phi_{\bar p} \phi_{p}} - \bar{v}^{\kappa\kappa}_{\phi_p \varphi_{\bar p} \phi_{ p} \phi_{\bar p}}
\right) \nonumber \\
&& \hspace{0.9cm}-  \left( \bar{v}^{\rho\rho}_{\varphi_p \phi_{\bar p} \phi_{p} \phi_{\bar p}} -
\bar{v}^{\rho\rho}_{\varphi_{\bar p} \phi_{p} \phi_{\bar p} \phi_{p}}  + \bar{v}^{\rho\rho}_{\phi_{\bar p}
\varphi_p \phi_{\bar p} \phi_{p}} - \bar{v}^{\rho\rho}_{\phi_p \varphi_{\bar p} \phi_{ p} \phi_{\bar p}}
\right)\Big] \, \bar Z_{p \bar p}
\end{eqnarray}
\end{widetext}
from the energy kernel, Eqs.~(\ref{kernel1}-\ref{kernel3}). The first two lines correspond to the spurious
self-pairing energy on the single-reference level, while the last four lines represent the additional self-pairing
energy contributing to multi-reference calculations.

\subsubsection{Chosen strategy}

The spurious terms causing divergences and steps, Eqs.~(\ref{eq:cor_rr}) and~(\ref{eq:cor_kk}), are the consequence of
a technical problem, i.e.\ using the GWT beyond its strict domain of validity. This could in principle be avoided by using
directly the standard Wick theorem to define the MR energy kernels, Eq.~(\ref{eneroverlap}). However, it is not
clear at present if this would be numerically feasible. While the direct evaluation of Eq.~(\ref{eneroverlap}) might
be too costly as it requires the explicit computation of the overlaps $\langle \Phi_0 | \Phi_1,\nu,\mu \rangle$
for all pairs of states $(\nu,\mu)$, calculating the correction, (\ref{eq:cor_rr}) and~(\ref{eq:cor_kk}), is not, and
should become a standard procedure in the future. As we demonstrate in Ref.~\cite{bender07x} for pure PNR calculations, correcting $\mathcal{E}_{GWT}[0,1]$ for divergences and finite steps indeed has a visible effect on
the binding energy even when not accidentally hitting a divergence. The correction fluctuates on the order of 1
MeV and leads to much smoother deformation energy curves.

From a practitioner's point of view, it is indeed mandatory to correct for divergences and finite steps as they
seriously compromise, even inhibit, meaningful MR calculations. Correcting for self-interaction and self-pairing
terms would also be desirable, either through explicit construction of self-interaction and self-pairing free
energy functionals, or through the correction of existing parameterizations of the functionals. Indeed, such
spurious contributions to the energy functional violate the Pauli principle and hence are always unphysical.

However, correcting for self-interaction and self-pairing terms, Eqs.~(\ref{selfint}) and~(\ref{selfpair}), is a
complex issue, both conceptually and computationally, i.e.\ it modifies the SR functional in such a way that the
equations of motion are computationally more difficult to solve. However, it is clear that the impact of
self-interaction and self-pairing on observables should be scrutinized. Existing studies in the context of HK-DFT
for electronic systems \cite{perdew81a,koch01,Ull00aDFT,Leg02aDFT,Ruz07aDFT,Vydrov07aDFT} demonstrate that it is
not an easy task and that it does not necessarily lead to an improvement of the performance of the functionals. As
can be seen from Eqs.~(\ref{selfint}) and~(\ref{selfpair}), self-interaction and self-pairing are a consequence of the
violation of the Pauli principle within a pair of conjugated states. Even when removing both from the energy
functional, there remain terms in the energy functional that break the particle exchange (anti)symmetry between
states that are not conjugated. Eventually, it should not be forgotten that there is a price to pay for the
enormous calculational simplification that EDF methods bring to the many-body problem. Loosing antisymmetry on the
level of the two-body density matrix is part of it, and how far it should be restored has to be determined as a
compromise between the required precision and the necessary computational cost.

\subsubsection{Fixing the freedom in the regularization procedure}

In any regularization procedure, the quantity that results from isolating and throwing away an infinity necessarily depends on explicit or implicit choices that determine its finite value. In the present case, two elements are essential to fix such a freedom.

First, and as explained above, we {\it choose} not to subtract finite self-interaction and self-pairing contributions which contaminate both SWT- and GWT-based expressions of MR EDF kernels. The second element relates to the basis used to proceed to the regularization. Although the GWT can be applied in any quasi-particle basis, which makes it so powerful, the application of the SWT necessitates to find a {\it specific} basis to express $| \Phi_1 \rangle$ in terms of $| \Phi_0 \rangle$, which we have managed to do as explained in section~\ref{sect:overlap:contra}. Of course, other bases exist that allow the application of the SWT. One\footnote{In fact, there is an infinity of such bases associated with the freedom to choose the third vacua with respect to which $| \Phi_0 \rangle$ and $| \Phi_1 \rangle$ must be expressed when using the technics of Refs.~\cite{donau98,Dob00}.} such basis~\cite{donau98,Dob00} relies on diagonalizing the matrix $Z^{0 \dagger} \, Z^{1}$ rather than on applying the BMZ decomposition to the Bogoliubov transformation connecting $| \Phi_0 \rangle$ and $| \Phi_1 \rangle$. In the resulting basis, the comparison between SWT- and GWT-based formulae for operators matrix elements works just as explained in section~\ref{sec:transdens} in the sense that terms proportional to $\langle \Phi_0 | \Phi_1, \nu \rangle \langle \Phi_0 | \Phi_1, \nu \rangle/ \langle \Phi_0 | \Phi_1  \rangle^{2}$ are present in the GWT-based formula but are absent in the SWT-based one. Removing those terms from the GWT-based kernel would again eliminate divergences and steps. However, this would lead to a different regularized energy kernel as the factors and wave-functions weighting such terms are not the same as in the basis we prioritize\footnote{One noticeable exception is PNR for which all convenient bases are the same.}. Consequently, the choice of the basis used to proceed to the regularization impacts the final regularized kernel. Going from one basis to another amounts to reshuffling finite spurious contributions between the different lines of Eqs.~\ref{eneroverlap} or~\ref{eneroverlapdirectGWT}.

Knowing that the regularization procedure is necessarily basis-dependent, just as standard self-interaction correction methods in DFT are~\cite{perdew81a}, the arguments that led us to prefer the basis we advocate are twofold. First, given $| \Phi_0 \rangle$ and $| \Phi_1 \rangle$, the application of the BMZ decomposition defines a unique basis, independently on the actual representation of the two states. On the contrary, using the basis that diagonalizes $Z^{0 \dagger} \, Z^{1}$ would require an extra argument to choose among the infinite number of vacua that can be used to express $| \Phi_0 \rangle$ and $| \Phi_1 \rangle$. Second, the basis we propose can always be found whereas it exists (rare) cases for which one cannot diagonalize the matrix $Z^{0 \dagger} \, Z^{1}$~\cite{Dob00}.

Eventually, the motivated choice of the basis we advocate and the decision to postpone to later the more challenging correction for self-interaction and self-pairing processes are the two elements used to fix the freedom that accompanies the present regularization procedure.

\subsection{Correcting higher-order functionals}
\label{sec:higherorder}

The cure proposed in the previous section to the problems faced when constructing a MR energy density functional
based on the GWT is specific to bilinear functionals. However, realistic functionals contain higher-order
dependencies on the density matrices associated with many-body correlations and three-body forces. Those
higher-order dependencies generate additional spurious contributions to the energy kernel $\mathcal{E}_{GWT}[0,1]$
which also have to be corrected for. The generalization of the strategy that we have followed is straightforward,
as long as one considers integer powers of the density matrices. Indeed, one can formally relate them to
higher-order multi-body forces in the Hamiltonian and identify, as we did for two-body forces, which terms are
zero when using the SWT and replaced by non-zero terms when using the GWT. However, working out the SWT becomes
lengthy and rather tedious for multi-body forces.

Thankfully, one does not need to apply the SWT explicitly, but rather proceed backwards using the connection
between the SWT and the GWT discussed in Sec.~\ref{sec:transdens}. This is more convenient because the GWT is
easy to apply to any Hamiltonian. For a $N$-body force, the SWT is recovered from the GWT by expressing $n$-body
transitions densities, with $n \in [0,N]$, through Eqs.~(\ref{lien1}-\ref{lien3}) and by using the identity
\begin{eqnarray}
\frac{ \langle \Phi_0   | \Phi_1, \nu_1 \rangle }
{\langle \Phi_0  | \Phi_1 \rangle} \cdots \frac{ \langle
\Phi_0   | \Phi_1, \nu_n \rangle } {\langle \Phi_0
 | \Phi_1 \rangle} = \frac{ \langle \Phi_0
| \Phi_1, \nu_1, \ldots,\nu_n \rangle } {\langle \Phi_0
 | \Phi_1 \rangle} \, , \label{eq:v1vn}
\end{eqnarray}
which generalizes in passing the notations introduced in Eqs.~(\ref{eq:define_nu}-\ref{eq:define_nu2}) in an obvious manner and which is valid as long as all $\nu_i$'s belong to different
conjugated pairs. When two or more $\nu_i$'s belong to the same
conjugated pair, the validity of Eq.~\ref{eq:v1vn} is lost and the left-hand side factor should be replaced by $\langle \Phi_0  | \Phi_1, \nu_1, \cdots,\nu_n \rangle \equiv 0$. Using such shortcuts, one can obtain an explicit "SWT" form for the matrix element of multi-body forces in a
very economical way.

Let us illustrate the above strategy for a three-body interaction. This will correspond to complementing the SR-EDF
$\mathcal{E}[\rho^{00}, \kappa^{00}, \kappa^{00 \, \ast}]$ and the energy kernel $\mathcal{E}_{GWT}[0,1]$ with a
trilinear component. The three-body interaction is written as
\begin{eqnarray}
\label{3bodyforce}
v_{123}
= \frac{1}{36} \sum_{ijklmn} \bar{v}_{ijklmn} \, a^+_i a^+_j a^+_k a_n a_m a_l \, ,
\end{eqnarray}
where the three-body force matrix elements are fully antisymmetric. Within the strict projected-GCM approximation,
the application of the GWT leads to
\begin{eqnarray}
\label{eq:wickthree}
\lefteqn{
E^{\rho\rho\rho}_{GWT}[0,1]+E^{\rho\kappa\kappa}_{GWT}[0,1]
}\nonumber \\
& = &   \tfrac{1}{6} \sum_{ijklmn} \bar{v}_{ijklmn} \,
        \rho^{01}_{li} \, \rho^{01}_{mj} \, \rho^{01}_{nk}
      \nonumber \\
&   & + \tfrac{1}{4} \sum_{ijklmn} \bar{v}_{ijklmn} \, \rho^{01}_{li} \,
        \kappa^{10^*}_{jk} \, \kappa^{01}_{mn} \, .
\end{eqnarray}
Operating as described above leads to the expression of the overlap (see Appendix~\ref{sec:threebody}) as it would
be obtained from Eq.~(\ref{eq:BCSqp}) using the SWT. Such an expression complements the last line of
Eq.~(\ref{eneroverlap}) for the two-body Hamiltonian.

As before, we now define the trilinear energy kernel $\mathcal{E}^{\rho\rho\rho}_{GWT}[0,1]$ and
$\mathcal{E}^{\rho\kappa\kappa}_{GWT}[0,1]$ within the EDF approach in close analogy to the two terms of
Eq.~(\ref{eq:wickthree}). The two contributions are expressed in terms of the effective vertices
$\bar{v}^{\rho\rho\rho}$ and $\bar{v}^{\rho\kappa\kappa}$, respectively. Just as for the bilinear parts,
$\mathcal{E}^{\rho\rho\rho}_{GWT}[0,1]$ and $\mathcal{E}^{\rho\kappa\kappa}_{GWT}[0,1]$ need now to be corrected for spurious
contributions. Proceeding in a similar way as for the bilinear functional, and using the SWT as a reference point,
Eq.~(\ref{eq:threedirect}), the terms to be removed are
\begin{widetext}
\begin{eqnarray}
\label{eq:cor_rrr} \mathcal{E}^{\rho \rho \rho}_{CG}[0,1] = \frac{1}{6}\sum_{p>0, \nu} && \Big(
\bar{v}^{\rho\rho\rho}_{\varphi_{\bar p} \varphi_{\bar p} \varphi_{\bar \nu}  \phi_{\bar p} \phi_{\bar p}
\varphi_{\bar \nu}}
-\bar{v}^{\rho\rho\rho}_{\varphi_{\bar p} \varphi_{p} \varphi_{\bar \nu}
 \phi_{\bar p} \phi_{p}
\varphi_{\bar \nu}} -\bar{v}^{\rho\rho\rho}_{\varphi_{p} \varphi_{\bar
p} \varphi_{\bar \nu}  \phi_{p} \phi_{\bar p} \varphi_{\bar \nu}} +
\bar{v}^{\rho\rho\rho}_{\varphi_{p} \varphi_{p} \varphi_{\bar \nu}
\phi_{p} \phi_{p} \varphi_{\bar \nu}} \nonumber \\ 
&& + \bar{v}^{\rho\rho\rho}_{\varphi_{\bar p} \varphi_{\bar \nu}
\varphi_{\bar p} \phi_{\bar p} \varphi_{{\bar \nu}} \phi_{\bar p}}
-\bar{v}^{\rho\rho\rho}_{\varphi_{p} \varphi_{\bar \nu} \varphi_{\bar p}
 \phi_{p} \varphi_{{\bar \nu}}
\phi_{\bar p}} -\bar{v}^{\rho\rho\rho}_{\varphi_{\bar p}
\varphi_{\bar \nu} \varphi_{p} \phi_{\bar p} \varphi_{{\bar \nu}} \phi_{p}} +
\bar{v}^{\rho\rho\rho}_{\varphi_{p} \varphi_{\bar \nu} \varphi_{p} \phi_{p} \varphi_{{\bar \nu}} \phi_{p}} \nonumber \\ 
&&+\bar{v}^{\rho\rho\rho}_{\varphi_{\bar \nu}\varphi_{\bar p}
\varphi_{\bar p} \varphi_{{\bar \nu}} \phi_{\bar p} \phi_{\bar p}}
-\bar{v}^{\rho\rho\rho}_{\varphi_{\bar \nu} \varphi_{\bar p} \varphi_{p}
 \varphi_{{\bar \nu}}\phi_{\bar
p}  \phi_{p}} -\bar{v}^{\rho\rho\rho}_{\varphi_{\bar \nu} \varphi_{p} \varphi_{\bar p}  \varphi_{{\bar \nu}}
\phi_{p} \phi_{\bar p}} + \bar{v}^{\rho\rho\rho}_{\varphi_{\bar \nu} \varphi_{p} \varphi_{p} \varphi_{{\bar \nu}}
\phi_{p}\phi_{p}} \Big) \;  \left(\bar Z_{ p \bar p }\right)^{2}
 \nonumber \\
+\frac{1}{6}\sum_{p>0, \nu} && \Big( \bar{v}^{\rho\rho\rho}_{ \varphi_{\bar p} \varphi_{\bar p} \varphi_{\bar
\nu} \phi_{\bar p} \phi_{\bar p} \phi_{{\bar \nu}}} -\bar{v}^{\rho\rho\rho}_{ \varphi_{\bar p} \varphi_{p}
\varphi_{\bar \nu}
 \phi_{\bar p} \phi_{p}
\phi_{\bar \nu}} -\bar{v}^{\rho\rho\rho}_{\varphi_{p} \varphi_{\bar p}
\varphi_{\bar \nu} \phi_{p} \phi_{\bar p} \phi_{\bar \nu}} +
\bar{v}^{\rho\rho\rho}_{\varphi_{p} \varphi_{p} \varphi_{\bar \nu}  \phi_p \phi_p \phi_{\bar \nu}} \nonumber \\ 
&& +\bar{v}^{\rho\rho\rho}_{\varphi_{\bar p} \varphi_{\bar \nu}
\varphi_{\bar p} \phi_{\bar p} \phi_{\bar \nu} \phi_{\bar p}}
-\bar{v}^{\rho\rho\rho}_{\varphi_{p} \varphi_{\bar \nu} \varphi_{\bar p}
 \phi_{p} \phi_{\bar \nu}
\phi_{\bar p}} -\bar{v}^{\rho\rho\rho}_{\varphi_{\bar p}
\varphi_{\bar \nu} \varphi_{p} \phi_{\bar p} \phi_{\bar \nu} \phi_{p}}+
\bar{v}^{\rho\rho\rho}_{\varphi_{p} \varphi_{\bar \nu} \varphi_{p} \phi_p \phi_{\bar \nu} \phi_p} \nonumber \\ 
&&+\bar{v}^{\rho\rho\rho}_{\varphi_{\bar \nu} \varphi_{\bar p}
\varphi_{\bar p} \phi_{\bar \nu}\phi_{\bar p} \phi_{\bar p}}
-\bar{v}^{\rho\rho\rho}_{\varphi_{\bar \nu} \varphi_{\bar p} \varphi_{p}
 \phi_{\bar \nu}\phi_{\bar p}
\phi_{p}}-\bar{v}^{\rho\rho\rho}_{\varphi_{\bar \nu} \varphi_{p}
\varphi_{\bar p} \phi_{\bar \nu}\phi_{p}  \phi_{\bar p}} +
\bar{v}^{\rho\rho\rho}_{\varphi_{\bar \nu} \varphi_{p}\varphi_{p}
\phi_{{\bar \nu}} \phi_p \phi_p}
 \Big) \;
 \left(\bar Z_{ p \bar p }\right)^{2} \; \bar{Z}_{\bar
\nu \nu}  \,  ,
\end{eqnarray}
and
\begin{eqnarray}
\label{eq:cor_kkr} \mathcal{E}^{\rho \kappa \kappa}_{CG}[0,1] = \frac{1}{4}\sum_{p>0, \nu} &&\Big(
\bar{v}^{\rho\kappa\kappa}_{\varphi_{\bar \nu} \varphi_{\bar p} \varphi_p \varphi_{\bar \nu} \phi_{\bar p}
\phi_{p}} - \bar{v}^{\rho\kappa\kappa}_{\varphi_{\bar \nu} \varphi_{\bar p} \varphi_p \varphi_{\bar \nu} \phi_{p}
\phi_{\bar p}} - \bar{v}^{\rho\kappa\kappa}_{\varphi_{\bar \nu} \varphi_{p} \varphi_{\bar p} \varphi_{\bar \nu}
\phi_{\bar p} \phi_{p}} + \bar{v}^{\rho\kappa\kappa}_{\varphi_{\bar \nu} \varphi_{p} \varphi_{\bar p}
\varphi_{\bar \nu} \phi_{p} \phi_{\bar p}}
\nonumber \\
&& +\bar{v}^{\rho\kappa\kappa}_{\varphi_{\bar p} \varphi_{\bar \nu}
\phi_{\nu} \phi_{\bar p} \phi_{\bar p} \phi_{p}}
-\bar{v}^{\rho\kappa\kappa}_{\varphi_{\bar p} \varphi_{\bar \nu}
\phi_{\nu} \phi_{\bar p} \phi_{p} \phi_{\bar p}}
-\bar{v}^{\rho\kappa\kappa}_{\varphi_{p} \varphi_{\bar \nu} \phi_{\nu}
\phi_{p} \phi_{\bar p} \phi_{p}}
+\bar{v}^{\rho\kappa\kappa}_{\varphi_{p} \varphi_{\bar \nu} \phi_{\nu}
 \phi_{p} \phi_{p} \phi_{\bar p}}
\nonumber \\
&&
   + \bar{v}^{\rho\kappa\kappa}_{\varphi_{\bar p} \varphi_{\bar p} \varphi_{p} \phi_{\bar p} \varphi_{{\bar \nu}} \phi_{\nu}}
   - \bar{v}^{\rho\kappa\kappa}_{\varphi_{\bar p} \varphi_p \varphi_{\bar p} \phi_{\bar p} \varphi_{\bar \nu} \phi_{\nu}}
   - \bar{v}^{\rho\kappa\kappa}_{\varphi_p \varphi_{\bar p} \varphi_{ p}  \phi_{p} \varphi_{\bar \nu} \phi_{\nu}}
   + \bar{v}^{\rho\kappa\kappa}_{\varphi_{p} \varphi_{p} \varphi_{\bar p} \phi_{p} \varphi_{\bar \nu} \phi_{\nu}} \Big) \;
     \left(\bar Z_{ p \bar p }\right)^{2}
\nonumber \\
+ \frac{1}{4}\sum_{p>0, \nu} && \Big( \bar{v}^{\rho\kappa\kappa}_{\varphi_{\bar p} \varphi_{\bar p} \varphi_{p}
\phi_{\bar p} \phi_{\bar \nu} \phi_{\nu}}
   - \bar{v}^{\rho\kappa\kappa}_{\varphi_{\bar p} \varphi_{p} \varphi_{\bar p} \phi_{\bar p} \phi_{\bar \nu} \phi_{\nu}}
   - \bar{v}^{\rho\kappa\kappa}_{\varphi_{p} \varphi_{\bar p} \varphi_{p} \phi_{p} \phi_{\bar \nu} \phi_{\nu}}
   + \bar{v}^{\rho\kappa\kappa}_{\varphi_{p} \varphi_{p} \varphi_{\bar p} \phi_{p} \phi_{\bar \nu} \phi_{\nu}}
\nonumber \\
&& + \bar{v}^{\rho\kappa\kappa}_{\varphi_{\bar p} \varphi_{\bar \nu} \varphi_{\nu}  \phi_{\bar p} \phi_{\bar p} \phi_{p}}
   - \bar{v}^{\rho\kappa\kappa}_{\varphi_{\bar p} \varphi_{\bar \nu} \varphi_{\nu}  \phi_{\bar p} \phi_{p} \phi_{\bar p}}
   - \bar{v}^{\rho\kappa\kappa}_{\varphi_{p} \varphi_{\bar \nu} \varphi_{\nu}  \phi_{p} \phi_{\bar p} \phi_{p}}
   + \bar{v}^{\rho\kappa\kappa}_{\varphi_{p} \varphi_{\bar \nu} \varphi_{\nu} \phi_{p} \phi_{p} \phi_{\bar p}}
\nonumber \\
&& +\bar{v}^{\rho\kappa\kappa}_{\varphi_{\bar \nu} \varphi_{\bar p} \varphi_{p} \phi_{\bar  \nu} \phi_{\bar p} \phi_{p}}
   - \bar{v}^{\rho\kappa\kappa}_{\varphi_{\bar \nu} \varphi_{\bar p} \varphi_{p} \phi_{\bar  \nu} \phi_{p} \phi_{\bar p}}
   - \bar{v}^{\rho\kappa\kappa}_{\varphi_{\bar \nu} \varphi_{p} \varphi_{\bar p} \phi_{\bar  \nu} \phi_{\bar p} \phi_{p}}
   + \bar{v}^{\rho\kappa\kappa}_{\varphi_{\bar \nu} \varphi_{p} \varphi_{\bar p} \phi_{\bar \nu} \phi_{p} \phi_{\bar p}}\Big) \;
     \left(\bar Z_{ p \bar p }\right)^{2} \; \bar{Z}_{\bar \nu \nu}   \, .
\end{eqnarray}
\end{widetext}
We have not made use of any antisymmetry properties of the vertices in Eqs.~(\ref{eq:cor_rrr}) and~(\ref{eq:cor_kkr}),
even though the vertex $\bar{v}^{\rho \kappa\kappa}$ can always be chosen to be antisymmetric, at least with
respect to the second and third indices on the one hand and to the fifth and sixth indices on the other.

A few observations similar to those made for the correction of the bilinear functional can be made for the
trilinear functional. First, $\mathcal{E}^{\rho \rho \rho}_{CG}[0,1]$ and $\mathcal{E}^{\rho \kappa
\kappa}_{CG}[0,1]$ sum up to zero if, and only if, both effective vertices refer to the same fully antisymmetric
three-body interaction. Second, both correction terms are zero for $| \Phi_0 \rangle = | \Phi_1 \rangle$ and do
not modify the underlying diagonal kernel $\mathcal{E}[0,0]$, keeping valid the link between the SR functional and
the MR one $\mathcal{E}[\rho,\kappa,\kappa^{\ast}] = \mathcal{E}[0,0]$. Third, the correction to the functional kernel
$\mathcal{E}_{GWT}[0,1]$ is independent of the normalization coefficient $\tilde{\mathcal{C}}_{01}$.

Note finally that the three-body terms should also be corrected for spurious self-interaction and self-pairing
problems. Expressions equivalent to~(\ref{selfint}) and~(\ref{selfpair}) can be derived without difficulty for
three-body vertices (not shown here).

\section{Application to Particle Number Restoration}
\label{PNR}

In the present section, we specify the previous findings to PNR calculations. A more extensive discussion of that
particular case, including results of realistic calculations, is presented in Ref.~\cite{bender07x}. When
considering one kind of particles only, the MR set appropriate to PNR is given by the ensemble of quasi-particle
vacua $\{| \Phi_{\varphi} \rangle ; \varphi \in [0,2\pi]\}$ which correspond to states rotated in gauge space by
an angle $\varphi$. The MR energy functional that amounts to restoring the particle number $N$ is given by
\begin{equation}
\label{scalar2}
\mathcal{E}^{N}
\equiv \int_{0}^{2\pi} \!\!\! d\varphi \, \frac{e^{-iN\varphi}}{2\pi \, c^{2}_{N}} \, \mathcal{E}[0,\varphi] \,
       \langle  \Phi_0 | \Phi_{\varphi} \rangle \,  ,
\end{equation}
with
\begin{equation}
\label{weight}
c^{2}_{N}
 = \int_{0}^{2\pi} d\varphi \, \frac{e^{-i N \varphi}}{2\pi} \, \langle  \Phi_0 | \Phi_{\varphi} \rangle  \, ,
\end{equation}
which is real.

In the previous equations, the state $| \Phi_0 \rangle$ is the quasi-particle vacuum at gauge angle $\varphi=0$.
Focusing on projection after variation, such a state is obtained from a self-consistent (possibly constrained) SR
calculation. Considering the case of an even-even nucleus, the vacuum state, possibly breaking time-reversal
invariance, is written in its canonical basis $\{a_{i}, a^{+}_{i}\}$ as
\begin{equation}
\label{eq:PNRphi0}
| \Phi_0 \rangle
= \prod_{p > 0} (u_p + v_p \, a^+_p a^+_{\bar p}) \,| 0 \rangle \, ,
\end{equation}
where $| 0 \rangle$ is the particle vacuum and where $\{u_{p};v_{p}\}$ are real numbers. According to our
convention, the product in Eq.~(\ref{eq:PNRphi0}) only runs over the "positive" half of the basis. The  state
$| \Phi_0 \rangle$ is normalized, with the convention that $u^{2}_{p}+v^{2}_{p}=1$. To underline the link with
previous sections, it is worth pointing out that Eq.~(\ref{eq:PNRphi0}) corresponds to a state
$| \Phi_0 \rangle = \mathcal{C}_0 \, \prod_\nu \alpha_\nu| 0 \rangle$, where the quasi-particle operators
$\{\alpha_{\nu}, \alpha^{+}_{\nu}\}$ are defined through a BCS-type transformation
\begin{equation}
\alpha^+_\nu
= U^0_{\nu \nu} a^+_\nu + V^0_{\bar \nu \nu} a_{\bar \nu}
\, ,
\end{equation}
with $\nu=p$ or $\bar p$ and where we have defined $v_p \equiv V^{0 \, \ast}_{p \bar p}= - V^{0 \, \ast}_{\bar p
p}$ and $u_p \equiv U^{0 \, \ast}_{p p}= U^{0 \, \ast}_{\bar p \bar p}$. The normalization of $| \Phi_{0} \rangle$
corresponds to $|\mathcal{C}_0|^{2} \, \prod_{p >0} v^{2}_{p} = 1$.

The second vacuum required to construct the energy kernel $\mathcal{E}[0,\varphi]$, i.e.\ where here $| \Phi_1 \rangle\equiv |
\Phi_{\varphi} \rangle$. This state is obtained by rotating $| \Phi_0 \rangle$ in gauge space by an angle $\varphi$, i.e.
\begin{equation}
\label{eq:transfo_phi}
| \Phi_{\varphi} \rangle
= e^{i \varphi \hat N} \, | \Phi_0 \rangle
= \prod_{p > 0} (u_p + v_p \, e^{2i\varphi} \, a^+_p a^+_{\bar p}) | 0 \rangle \, .
\end{equation}
From a technical point of view, particle-number restoration is simple because all vacua belonging to the MR set
share the same canonical basis. As a result, the Bogoliubov transformation linking any pair of vacua in the set is
itself canonical, i.e.
\begin{equation}
\label{transfocanoPNR}
\beta_\nu \equiv e^{i \varphi \hat N} \, \alpha_\nu \, e^{-i \varphi \hat N}
= \bar A^*_{\nu \nu} \alpha_{\nu} + \bar B^*_{\bar \nu \nu} \alpha^+_{\nu} \, ,
\end{equation}
with
\begin{eqnarray}
\bar A^*_{\bar p \bar p}
& = & \bar A^*_{p p}
  =   e^{-i\varphi} \left( u_p^2 + v_{p}^2 e^{2i\varphi} \right) \, ,
      \\
\bar B^*_{p \bar p}
& = & -\bar B^*_{\bar p p}
  =   u_{p} v_{p} (e^{i\varphi} -e^{-i\varphi}) \, .
\end{eqnarray}
It is also interesting to compute the transition densities through Eq.~(\ref{eq:rho01rho0}). For example, it leads
for the diagonal normal transition density matrix
\begin{equation}
\label{eq:rhophi}
\rho^{0\varphi}_{pp}
=   V^{0 \, \ast}_{p \bar p} \, V^{0}_{p \bar p}
  + U^{0}_{pp} \, \frac{\bar B^*_{p\bar p}}{\bar A^*_{\bar p \bar p }}
       \, V^{0}_{p \bar p}
= \frac{v^2_p \, e^{2i\varphi}}
       {u^2_p + v^2_{ p} \, e^{2i\varphi}} \, ,
\end{equation}
whereas the anomalous ones are obtained in the same way as
\begin{eqnarray}
\label{eq:kap1phi} \kappa^{0\varphi}_{p\bar p}
& = & \frac{u_{p} v_{p} e^{2 i \varphi}}
           {u_{p}^2 + v_{p}^2 \, e^{2 i \varphi}}
     \, , \\
\label{eq:kap0phi}
\kappa^{\varphi 0 \, \ast}_{p\bar p}
& = &  \frac{u_{p} v_{p}}
            {u_{p}^2 + v_{p}^2 \, e^{2 i \varphi}}
\, .
\end{eqnarray}
Before turning to the correction, let us discuss the relative phase between the two vacua. On the one hand,
applying Eq.~(\ref{eq:overc}) provides
\begin{equation}
\langle \Phi_0 | \Phi_{\varphi} \rangle
= \mathcal{C}_{01} \, \prod_{p > 0} e^{-i\varphi} \, \big( u_p^2 + v_{p}^2 e^{2i\varphi} \big) \, ,
\end{equation}
whereas on the other hand, the explicit canonical forms given by Eqs.~(\ref{eq:PNRphi0}-\ref{eq:transfo_phi}) lead
to $\langle \Phi_0  | \Phi_{\varphi} \rangle = \prod_{p > 0} \left( u_p^2 + v_{p}^2 e^{2i\varphi} \right)$. Thus,
the phase convention associated with the canonical forms is equivalent to having chosen $\mathcal{C}_{01} \prod_{p >
0} e^{-i\varphi} = 1$.

The fact that the Bogoliubov transformation $(A,B)$ is canonical from the outset is a specificity of PNR. In
particular, it corresponds to having the trivial unitary transformations $C=D=1$ in Eq.~(\ref{eq:cano}). For almost
all other cases of interest, however, i.e.\ Angular Momentum Projection and/or GCM-type Multi-Reference
calculations, $\bar A$ and $\bar B$ cannot be obtained analytically. This means that the matrices $A$ and $B$ must
be computed using Eq.~(\ref{eq:uv2}) for each pair of vacua belonging to the MR set. To this end, methods that will
be outlined in Sec.~\ref{sectionzeropairing} have to be applied. Their implementation is underway and will be
discussed elsewhere.

Finally, the upper and lower components associated with the quasi-particle operators $\{\tilde \alpha_{p}, \tilde
\alpha^{+}_{p}\}$ introduced through Eqs.~(\ref{eq:def_sp2}-\ref{eq:def_sp1}) take the form
\begin{alignat}{3}
| \varphi_{\bar{p}} \rangle & = -v_p \, | \bar{p} \rangle \, , & \qquad
| \varphi_p \rangle         & =  v_p \, | p \rangle \, , \nonumber \\
| \phi_p \rangle            & = \phantom{-} u_p \, | p \rangle \, , &
| \phi_{\bar{p}} \rangle    & =  u_p \, | \bar{p} \rangle \, ,
\end{alignat}
where $| p \rangle$ and $| \bar{p} \rangle$ denote the single-particle states created respectively by $a^+_p$ and
$a^+_{\bar{p}}$, respectively.

\subsubsection{Bilinear functional}

The different types of spurious contributions to the bilinear energy kernel read in the PNR case as
\begin{widetext}
\begin{eqnarray}
\label{correctPNRbil1}
\mathcal{E}^{\rho \rho}_{CG}[0,\varphi]
&=& \frac{1}{2} \sum_{p>0}
    \Big(   \bar{v}^{\rho\rho}_{pppp}
          + \bar{v}^{\rho\rho}_{\bar{p}\bar{p}\bar{p}\bar{p}}
          + \bar{v}^{\rho\rho}_{p\bar{p}p\bar{p}}
          + \bar{v}^{\rho\rho}_{\bar{p}p\bar{p}p}
    \Big) \, (u_p v_p)^4 \,
    \frac{(e^{2i\varphi}-1)^2}{\left( {u_p}^2 + {v_{ p}}^2 e^{2i\varphi} \right)^2} \, , \\
&& \nonumber \\
\label{correctPNRbil2}
\mathcal{E}^{\kappa \kappa}_{CG}[0,\varphi]
&=& - \sum_{p>0} \bar{v}^{\kappa\kappa}_{p\bar{p}p\bar{p}} \, (u_p v_p)^4 \,
      \frac{(e^{2i\varphi}-1)^2}{\left( {u_p}^2 + {v_{ p}}^2 e^{2i\varphi} \right)^2} \, ,
      \\
\label{correctPNRbil3}
\mathcal{E}^{\rho \rho}_{SI}[0,\varphi]
&= & \frac{1}{2}\sum_{p>0} \Big( \bar{v}^{\rho\rho}_{pppp} + \bar{v}^{\rho\rho}_{\bar{p}\bar{p}\bar{p}\bar{p}} \Big)
     \Big[ v^4_p +  2 \, u^2_p v^4_p \, \frac{(e^{2i\varphi}-1)}{{u_p}^2 + {v_{ p}}^2 e^{2i\varphi}} \Big] \, ,
\\
\label{correctPNRbil4}
\mathcal{E}^{\kappa \kappa}_{SP}[0,\varphi]
& = & \sum_{p>0} \Big[ \bar{v}^{\kappa \kappa}_{p \bar pp \bar p}
                      - \frac{1}{2} \big(\bar{v}^{\rho\rho}_{p\bar{p}p\bar{p}} +\bar{v}^{\rho\rho}_{\bar{p}p\bar{p}p} \big)
                 \Big] \,
                 \Big[ (u_p v_p)^{2}  + (u^4_p v^2_p -u^2_p v^4_p) \frac{(e^{2i\varphi}-1)}{{u_p}^2 + {v_{ p}}^2 e^{2i\varphi}}\Big] \, .
\end{eqnarray}
Subtracting $\mathcal{E}^{\rho \rho}_{CG}[0,\varphi]$ and $\mathcal{E}^{\kappa \kappa}_{CG}[0,\varphi]$ from the PNR
energy kernel does not call for a modification of the underlying SR functional whereas removing $\mathcal{E}^{\rho
\rho}_{SI}[0,\varphi]$ and $\mathcal{E}^{\kappa \kappa}_{SP}[0,\varphi]$ would.

\subsubsection{Trilinear functional}

The spurious contribution to be removed from the GWT-based EDF kernel $\mathcal{E}^{\rho\rho\rho}[0,\varphi,\varphi']$ is given by
\begin{eqnarray}
\label{eq:cor_rrrPNR} \mathcal{E}^{\rho \rho \rho}_{CG}[0,\varphi,\varphi'] & = & \frac{1}{6}\sum_{p>0, \nu}
\Big\{\bar{v}^{\rho\rho\rho}_{{\bar p} {\bar p} \nu {\bar p} {\bar p} {\nu}} + \bar{v}^{\rho\rho\rho}_{{\bar p}
{p} \nu {\bar p} {p} {\nu}}+\bar{v}^{\rho\rho\rho}_{p {\bar p} \nu p {\bar p} {\nu}} + \bar{v}^{\rho\rho\rho}_{pp
\nu pp {\nu}} + \bar{v}^{\rho\rho\rho}_{{\bar p} \nu {\bar p} {\bar p} {\nu} {\bar p}}+
\bar{v}^{\rho\rho\rho}_{{p} \nu {\bar p}
 {p} {\nu} {\bar p}} \nonumber \\
&& +\bar{v}^{\rho\rho\rho}_{{\bar p} \nu {p}  {\bar p} {\nu} {p}} + \bar{v}^{\rho\rho\rho}_{ {p} \nu {p} {p} {\nu}
{p}}+\bar{v}^{\rho\rho\rho}_{\nu {\bar p} {\bar p}  {\nu} {\bar p} {\bar p}} +\bar{v}^{\rho\rho\rho}_{\nu {\bar p}
{p}
 {\nu} {\bar p} {p}}
+\bar{v}^{\rho\rho\rho}_{\nu {p}{\bar p} {\nu}  {p} {\bar p}} + \bar{v}^{\rho\rho\rho}_{\nu {p} {p} {\nu}  {p}
{p}} \Big\} \,  (u_p v_p)^4 \frac{(1-e^{2i\varphi})^2}{\left( {u_p}^2 + {v_{ p}}^2 e^{2i\varphi} \right)^2} \;
\rho^{0\varphi '}_{\nu \nu} \, .
\end{eqnarray}
\end{widetext}
In this expression appears a transition density matrix as defined by Eq.~(\ref{eq:rhophi}) and evaluated at a second
gauge angle $\varphi '$. This second gauge angle refers to the fact that the isospin degree of freedom must be
treated explicitly when correcting the trilinear functional, as there will be terms that are bilinear in one
isospin and linear in the other. As a result, both projections on neutron and proton numbers must be considered
explicitly. The isospin could be omitted for the bilinear functional because, as long as we do not mix neutron and
proton in the mean-field and only deal with neutron-neutron and proton-proton pairing, the conjugated states $p$
and $\bar p$ always refer to the same isospin and are rotated by the same angle $\varphi$ in gauge space. In
$\mathcal{E}^{\rho\rho\rho}$ and $\mathcal{E}^{\rho\kappa\kappa}$, however, the particle $\nu$ {may} have a different
isospin from the one of the pair $(p,\bar p)$ and a second gauge angle $\varphi '$ must be attached to it.

Finally, the contribution to be removed from the functional energy kernel $\mathcal{E}^{\rho \kappa \kappa}$ reads as
\begin{widetext}
\begin{eqnarray}
\mathcal{E}^{\rho \kappa \kappa}_{CG}[0,\varphi,\varphi']
&=& \frac{1}{4}\sum_{p>0 \nu}
    \Big(  \bar{v}^{\rho\kappa\kappa}_{ \nu {\bar p}  p {\nu} {p}{\bar p}}
         - \bar{v}^{\rho\kappa\kappa}_{\nu {\bar p} p {\nu} {\bar p} {p}}
         + \bar{v}^{\rho\kappa\kappa}_{ \nu {p} {\bar p} {\nu} {\bar p} {p}}
         - \bar{v}^{\rho\kappa\kappa}_{\bar  {p} {\bar p}  {\nu} {p} {\bar p}}
    \Big) (u_p v_p)^4
    \frac{(1-e^{2i\varphi})^2}{\left( {u_p}^2 + {v_{ p}}^2 e^{2i\varphi} \right)^2} \rho^{0\varphi '}_{ \nu  \nu}
\nonumber \\
&+& \frac{1}{4}\sum_{p>0 \nu}
    \Big(   \bar{v}^{\rho\kappa\kappa}_{\bar p {\nu} \bar \nu  {\bar p} {p} {\bar p}}
          - \bar{v}^{\rho\kappa\kappa}_{\bar p  \nu {\bar \nu} {\bar p} {\bar p} {p}}
          - \bar{v}^{\rho\kappa\kappa}_{ {p} \nu {\bar\nu} {p} {\bar p} {p}}
          + \bar{v}^{\rho\kappa\kappa}_{{p} \nu { \bar\nu} {p} {p} {\bar p}}
    \Big) \, u^2_p (u_p v_p)^3
    \frac{(1-e^{2i\varphi})^2}{\left( {u_p}^2 + {v_{ p}}^2 e^{2i\varphi} \right)^2}\kappa^{\varphi ' 0 \, \ast}_{\nu \bar \nu}
\nonumber \\
&+& \frac{1}{4}\sum_{p>0 \nu}
    \Big(   \bar{v}^{\rho\kappa\kappa}_{\bar p {\bar p} {p} {\bar p} \nu \bar {\nu}}
          - \bar{v}^{\rho\kappa\kappa}_{\bar p p \bar p {\bar p}  {\nu} \bar{\nu}}
          + \bar{v}^{\rho\kappa\kappa}_{p \bar p p {p} {\nu} \bar {\nu}}
          - \bar{v}^{\rho\kappa\kappa}_{p p \bar {p} {p}  {\nu} \bar{\nu}}
    \Big) v^2_p (u_p v_p)^3
    \frac{(1-e^{2i\varphi})^2}{\left( {u_p}^2 + {v_{ p}}^2 e^{2i\varphi} \right)^2} \kappa^{0\varphi ' }_{\nu \bar \nu} \, ,
\end{eqnarray}
\end{widetext}
where, again, $\varphi' =\varphi$ if and only if $\nu$ and $p$ have the same isospin; otherwise $\varphi ' \neq
\varphi$.

\section{No-pairing case}
\label{sectionzeropairing}

Multi-reference EDF calculations are sometimes performed without an explicit account for pairing correlations;
i.e.\ the reference states of the MR set take the form of Slater determinants rather than quasi-particle vacua. We concentrate on such a case in the present section.
This is of particular interest since recent MR calculations based on triaxial cranked Slater
determinants and aiming at restoring angular momentum displayed divergences \cite{doba06a} which we believe
to be related to the problems discussed in the present work.

To investigate the zero-pairing realization of the formalism outlined above we consider a MR energy kernel associated
with two Slater determinants $\left| \Phi_0 \right\rangle$ and $\left| \Phi_1 \right\rangle$ given by
\begin{eqnarray}
\left| \Phi_0 \right\rangle
& = & \prod^N_{i=1} a^+_{i} \left| 0 \right\rangle \, ,
      \nonumber \\
\left| \Phi_1 \right\rangle
&=& \prod^N_{i=1} b^+_{i} \left| 0 \right\rangle \, ,
\end{eqnarray}
where $N$ represents the number of particles whereas $\{a^+_{i}\}_{i=1,N}$ and $\{b^+_{i}\}_{i=1,N}$ are the
creation operators associated with occupied, i.e. hole, single-particle states $\{\left| a_i \right\rangle\}_{i=1,N}$ and $\{\left| b_i
\right\rangle\}_{i=1,N}$, respectively. In the general case, we have expressed the two quasi-particle vacua,
Eq.~(\ref{eq:bogo0}), in a common, arbitrary single-particle basis. However, numerical applications often make use
of two non-equivalent sets of single-particle states to define the two vacua. This is naturally the case
when dealing with Slater determinants. Thus, and as a by-product, the following discussion of the zero-pairing
case will sketch how to proceed when the quasi-particle vacua are defined with respect to different
single-particle bases.

An important result of the present work (Section~\ref{illdefined}) consists in finding a convenient basis through
the BMZ decomposition of the Bogoliubov transformation connecting two quasi-particle vacua, Eq.~(\ref{eq:bcslike}).
The Slater determinant limit discussed in the present section offers a chance to illustrate
more intuitively how this works.

The single-particle subspace spanned by the $\{\left| b_i \right\rangle\}_{i=1,N}$ is a priori different from the one spanned by
the $\{\left| a_i \right\rangle\}_{i=1,N}$. However, it is clear that the subspace spanned by the addition of the
two bases is at maximum of dimension $2N$. Therefore, one can always complete the set of hole states of $\left|
\Phi_0 \right\rangle$ by an appropriate choice of $N$ particle states, denoted by $\{\left| a_{\bar k}
\right\rangle\}_{k=1,N}$, in such a way that
\begin{equation}
| b_i \rangle
=   \sum^N_{j=1} | a_j \rangle \langle a_j | b_i \rangle
  + \sum_{k = 1}^N | a_{\bar k} \rangle \, \langle a_{\bar k} | b_i \rangle
\, . \label{eq:phoriginal}
\end{equation}
Similarly, one can introduce a set of $N$ particle states for $\left| \Phi_1 \right\rangle$, denoted by $\{\left|
b_{\bar k} \right\rangle\}_{k=1,N}$  such that
\begin{equation}
| a_i \rangle
=   \sum^N_{j=1} | b_j \rangle \langle b_j | a_i \rangle
  + \sum^N_{ k = 1} | b_{\bar k} \rangle \, \langle b_{\bar k} | a_i \rangle
\, .  \label{eq:ph}
\end{equation}
In practice, the most convenient choice is to associate to each state $\left| b_i \right\rangle$ an intermediate
state $\left| d_i \right\rangle$ given by
\begin{equation}
| d_i \rangle \equiv | b_i \rangle - \sum^N_{j=1} | a_j \rangle \,
       \langle a_j | b_i \rangle \, .
\end{equation}
Then, the $N$ states $| a_{\bar k} \rangle$ are obtained by a Schmidt orthogonalization of the $N$ states $| d_i
\rangle$. If the state $| b_i \rangle$ can be completely described by the hole states of $| \Phi_0 \rangle$, the
associated $| d_i \rangle$ cancels out and the dimensionality of the considered matrix can be reduced.

We associate the creation operators $a^+_{\bar k}$ and $b^+_{\bar k}$ to the states $\left| a_{\bar k}
\right\rangle$ and $\left| b_{\bar k} \right\rangle$, respectively. One has in particular $a_{\bar k}\left| \Phi_0
\right\rangle = 0$ and $b_{\bar k}\left| \Phi_1 \right\rangle = 0$, for all $k =1, \ldots, N$.

Starting from the truncated space as introduced above, a formalism adapted to MR-EDF calculations with Slater
determinants could be introduced without referring to quasi particles at all. Then, expressions of the functional
kernels and the corrections to spurious processes could be obtained directly in the language of particles.
However, our present goal is to demonstrate how the formalism written previously in terms of quasi particles can
be directly applied to the particular case where pairing is not considered explicitly. The first step is thus to match the
particle creation/annihilation operators with quasi-particle creation/annihilation operators for Slater
determinants. Although this can be found in textbooks \cite{ring80a}, we present it below for the sake of a
self-contained description of our method.

\subsection{From particles to quasi particles}

The matching appropriate to the notation used in previous sections is achieved using the standard quasi-particle
representation of Slater determinants. We introduce the two sets of quasi-particle annihilation operators
$\alpha_{\nu}$ and $\beta_{\nu}$ and restart from the general form of the transformations between particles and
quasi-particles operators:
\begin{equation}
\label{eq:alpha_a}
\left( \begin{array}{c} \alpha \\ \alpha^+ \end{array} \right)
\equiv \left( \begin{array}{cc}
       U^{0^+} & V^{0^+} \\
       V^{0^T} & U^{0^T}
       \end{array} \right)_{4N}
       \left( \begin{array}{c} a \\ a^+ \end{array} \right) \, ,
\end{equation}
and
\begin{equation}
\label{eq:beta_b}
\left( \begin{array}{c} \beta \\ \beta^+ \end{array} \right)
\equiv \left( \begin{array}{cc}
       U^{1^+} & V^{1^+} \\
       V^{1^T} & U^{1^T}
       \end{array} \right)_{4N}
       \left( \begin{array}{c} b \\ b^+ \end{array} \right) \, .
\end{equation}
Here, the index $4N$ is used to stress that only a specific finite number of quasi-particle states is necessary.
The annihilation operators verify $\alpha_\nu \left| \Phi_0 \right\rangle = 0$ and $\beta_\nu \left| \Phi_1
\right\rangle = 0$ for $\nu=1, \ldots, 2N$. It is clear that this relation is fulfilled if $\alpha_\nu$ (resp.
$\beta_\nu$) is proportional to a linear combination of hole creation operators of $\left| \Phi_0 \right\rangle$
(resp. $\left| \Phi_1 \right\rangle$) or to a combination of its particle annihilation operators. This leads to a
simple block structure\footnote{It is worth mentioning that there is a flexibility in choosing the lower-right
block and upper-left blocks of $U^{0/1}$ and $V^{0/1}$ respectively. Here we choose the simplest prescription
which in our opinion is also the most convenient in applications.} for the matrices $U^{0/1}$ and $V^{0/1}$
\begin{equation}
U^{0/1}
\equiv \left( \begin{array}{cc}
       0 & 0 \\
       0 & 1
       \end{array} \right)_{2N}, \qquad
V^{0/1}
\equiv \left( \begin{array}{cc}
       1 & 0 \\
       0 & 0
       \end{array} \right)_{2N} \, ,
\end{equation}
which is obtained thanks to a specific ordering of hole and particle states, that is, we have
$\{a^+_\nu\}_{\nu=1,2N} \equiv (\{a^+_i \}_{i=1,N},\{a^+_{\bar \imath} \}_{ \imath=1,N})$.

\subsection{Bogoliubov transformation between two Slater determinants}

To express the matrices $A$ and $B$ associated with the transformation between the two sets of quasi-particles
$\{\alpha_{\nu}, \alpha^{\dagger}_{\nu}\}$ and $\{\beta_{\mu}, \beta^{\dagger}_{\mu}\}$ introduced above, we
first define the transformation
\begin{equation}
\label{eq:b_a}
\left( \begin{array}{c} b \\ b^+ \end{array} \right)
= \left( \begin{array} {cc}
  R & 0 \\
  0 & R^*
  \end{array} \right)_{4N}
  \left( \begin{array}{c} a \\ a^+ \end{array} \right) \, ,
\end{equation}
where $R$ is the overlap matrix that can be schematically written as
\begin{eqnarray}
R&\equiv& \left(
\begin{array} {cc}
\{ \left\langle  b_i \left.  \right| a_j \right\rangle \} &
\{\left\langle  b_i \left.  \right| a_{\bar{\jmath}} \right\rangle \}  \\
\{ \left\langle  b_{\bar{\imath}} \left.  \right| a_{j} \right\rangle  \} & \{ \left\langle  b_{\bar{\imath}}
\left. \right| a_{\bar{\jmath}} \right\rangle \}
\end{array}
\right)_{2N}
\equiv \left(
\begin{array} {cc}
\mathcal{A}^T & \mathcal{B}^{T} \\
\mathcal{X}^+ & \mathcal{Y}^+
\end{array}
\right)_{2N} \, . \label{eq:rrslater}
\end{eqnarray}
The matrices $A$ and $B$ can finally be deduced using Eqs.~(\ref{eq:alpha_a}), (\ref{eq:beta_b}) and
(\ref{eq:b_a})
\begin{eqnarray}
\label{eq:bogo:slater}
A &=& U^{0+} R^+ U^1 + V^{0+} R^T V^1 \, ,\nonumber \\
B &=& V^{0^T} R^+ U^{1}   + U^{0^T} R^T V^{1} \, ,
\end{eqnarray}
which, using the simple form of $U^{0/1}$ and $V^{0/1}$ leads to
\begin{eqnarray}
A &=&\left(
\begin{array} {cc}
\mathcal{A}  & 0 \\
0 &  \mathcal{Y}
\end{array}
\right)_{2N}, \qquad
B =\left(
\begin{array} {cc}
 0 & \mathcal{X} \\
\mathcal{B}
& 0
\end{array}
\right)_{2N} \, .
\end{eqnarray}

\subsection{Bloch-Messiah-Zumino decomposition}

The matrices $A$ and $B$ fulfill all usual relations associated with Bogoliubov transformations. Guided by the
standard BMZ theorem, we introduce the two matrices $S$ and $T$ defined as
\begin{eqnarray}
S & \equiv &  B^* B^T \,, \nonumber \\
T & \equiv & -AB^+ = B^* A^T  \, .
\end{eqnarray}
These two matrices play a role similar to that of the normal and anomalous density matrices in the SR case. In
particular, we have the following relationships
\begin{eqnarray}
\label{eq:st}
S^2 -S
& = & -TT^+  \,  ,
      \nonumber \\
S T
& = & T S^* \, .
\end{eqnarray}

However, it should be noted that the matrices $S$ and $T$ are \emph{not} related to the transition density matrices in any
trivial way. Using the expression of $A$ and $B$, $S$ can be written as
\begin{eqnarray}
S &=& \left(
\begin{array} {cc}
\mathcal{X}^*\mathcal{X}^T   & 0 \\
0 &  \mathcal{B}^*\mathcal{B}^T
\end{array}
\right)_{2N}
\equiv \left(
\begin{array} {cc}
\mathcal{S}^X   & 0 \\
0 &  \mathcal{S}^B
\end{array}
\right)_{2N} \, , \label{eq:s}
\end{eqnarray}
while $T$ takes the form
\begin{eqnarray}
T &=&  \left(
\begin{array} {cc}
0  & -\mathcal{A} \mathcal{B}^+ \\
+ (\mathcal{A}\mathcal{B}^+)^T &  0
\end{array}
\right)_{2N}\equiv \left(
\begin{array} {cc}
0  & \mathcal{T}  \\
-\mathcal{T}^T  &  0
\end{array}
\right)_{2N} \, ,  \label{eq:t}
\end{eqnarray}
in such a way that $T$ is antisymmetric as expected. In the following, it will be useful to have a more
explicit form of the matrices $\mathcal{S}^X$, $\mathcal{S}^B$ and $\mathcal{T}$. Using Eqs.~(\ref{eq:rrslater}) and
(\ref{eq:phoriginal}-\ref{eq:ph}) we deduce
\begin{eqnarray}
\label{eq:ssk} \mathcal{S}^X_{ij} & = & +\langle a_i | (1-\rho^{11}) | a_j \rangle^* \,  ,
      \nonumber \\
\mathcal{S}^B_{\bar{\imath} \bar{\jmath}} & = & +\langle a_{\bar{\imath}} | \rho^{11} | a_{\bar{\jmath}} \rangle \, ,
      \nonumber \\
\mathcal{T}_{i\bar{\jmath}} & = & -\langle a_i | \rho^{11} | a_{\bar{\jmath}} \rangle^* \, .
\end{eqnarray}
One can see from Eq.~(\ref{eq:ssk}) that all the information relative to the overlaps between the single-particle
wave-functions of the two bases is encoded into the three matrices $\mathcal{S}^X$, $\mathcal{S}^B$ and $\mathcal{T}$.

The first step to obtain a BMZ decomposition of $A$ and $B$ is to diagonalize the matrix $S$. The block diagonal
form of $S$, Eq.~(\ref{eq:s}), shows that the unitary transformation allowing to do so is also block diagonal; i.e.
diagonalizing $S$ is achieved by diagonalizing $\mathcal{S}^X$ in the subspace of hole states of $\left| \Phi_0
\right\rangle$ and $\mathcal{S}^B$ in the subspace of particle states of $| \Phi_0 \rangle$. As a result of such a
transformation, new hole and particle states are obtained for $| \Phi_0 \rangle$, denoted by $| \tilde a_i
\rangle$ and $| \tilde a_{\bar{\imath}} \rangle$, respectively. We define the corresponding creation and
annihilation operators by $\tilde a^+_i$ and $\tilde a_{\bar{\imath}}$ as well as the eigenvalues of $\mathcal{S}^X$ and $\mathcal{S}^B$ by $\lambda^x_i$
and $\lambda^b_{\bar{\imath}}$, respectively. Expressing the two relationships of Eq.~(\ref{eq:st}) in the
basis where $S$ is diagonal leads to
\begin{eqnarray}
\lambda^{b}_{\bar{\jmath}} (\lambda^{b}_{\bar{\jmath}} - 1) &=&  -\sum_{i=1}^{N} |\widetilde{\mathcal{T}}_{i
\bar{\jmath}}|^2 \, \, ,\\
&&\nonumber\\
\lambda^x_{i} (\lambda^x_{i} - 1) &=& -\sum_{\bar{\jmath}=1}^{N} |\widetilde{\mathcal{T}}_{i \bar{\jmath}}|^2 \, \, ,
\label{truc1}\\
&& \nonumber \\
\widetilde{\mathcal{T}}_{i \bar{\jmath}} \, (\lambda^x_{i}- \lambda^{b}_{\bar{\jmath}}) &=& 0
\,  \, ,
\end{eqnarray}
which shows that $\widetilde{\mathcal{T}}_{i \bar{\jmath}}$ may differ from zero only if
$\lambda^x_{i}= \lambda^{b}_{\bar j}$. Let us now consider the different possible
eigenvalues of $\mathcal{S}^{X}$. There exist three different cases
\begin{itemize}
\item $\lambda^x_{i} = 0$: this means that the eigenstate $| \tilde a_i \rangle$ of
$\mathcal{S}^X$ is orthogonal to all particle states of $\left| \Phi_1 \right\rangle$ and therefore can be written as
a linear combination of the occupied states in $\left| \Phi_1 \right\rangle$. Eq.~(\ref{truc1}) automatically
implies that $\widetilde{\mathcal{T}}_{i \bar{\jmath}}=0$ for all $\jmath=1, \ldots, N$.
\item $\lambda^x_{i} = 1$:
this means that the eigenstate $\left| \tilde a_i \right\rangle$ is fully contained in the subspace of particle
states of $\left| \Phi_1 \right\rangle$. Again, Eq.~(\ref{truc1}) implies that
$\widetilde{\mathcal{T}}_{i\bar{\jmath}}=0$ for all $\jmath=1, \ldots, N$.
\item $1<\lambda^x_{i} <0$: the corresponding eigenstate is neither entirely contained
in the space of particle states of $\left| \Phi_1 \right\rangle$ nor in the space of its hole states. This also
means that at least one matrix element $\widetilde{\mathcal{T}}_{i \bar{\jmath}}$ is non vanishing, for
$\jmath=1,\ldots,N$. Accordingly, there exists at least one eigenvalue $\lambda^{b}_{\bar{\jmath}}$ such that $\lambda^{b}_{\bar{\jmath}}=\lambda^x_{i}$.
\end{itemize}
The same classification could be made for the eigenvalues $\lambda^{b}_{\bar{\imath}}$ except that
$\lambda^{b}_{\bar{\imath}} = 0$ or $\lambda^{b}_{\bar{\imath}} = 1$ correspond in this case to eigenstates $| \tilde a_{\bar{\imath}} \rangle$ belonging to
particle or hole states of $| \Phi_1 \rangle$, respectively.

It could be checked that the diagonalization of $S$ is not affected by a unitary transformation that acts
separately among the particle or hole states of $\left| \Phi_1 \right\rangle$. Indeed, $\rho^{11}$ and
$(1-\rho^{11})$ entering Eq.~(\ref{eq:ssk}) are invariant under such a unitary transformation, respectively. To
advance towards the BMZ decomposition the Bogoliubov transformation~(\ref{eq:bogo:slater}), one now needs to
transform the (quasi-)particle states associated with $| \Phi_1 \rangle$. This can be achieved by repeating the above procedure for the
new matrices $S' \equiv B^T B^*$ and $T' \equiv B^T A$
\begin{eqnarray}
S' = \left(
\begin{array} {cc}
\mathcal{B}^T \mathcal{B}^*   & 0 \\
0 &  \mathcal{X}^T\mathcal{X}^*
\end{array}
\right)_{2N} =
\left(
\begin{array} {cc}
{\mathcal{S}'}^B  & 0 \\
0 &  {\mathcal{S}'}^X
\end{array}
\right)_{2N} \, ,
\end{eqnarray}
and
\begin{eqnarray}
T' &=& \left(
\begin{array} {cc}
0  & \mathcal{B}^T\mathcal{Y}  \\
-(\mathcal{B}^T\mathcal{Y})^T &  0
\end{array}
\right)_{2N} = \left(
\begin{array} {cc}
0  & \mathcal{T}'  \\
-{\mathcal{T}'}^T  &  0
\end{array}
\right)_{2N} \, ,
\end{eqnarray}
where the matrix elements now read
\begin{eqnarray}
{\mathcal{S}'}^B_{ij}
&=& +\left\langle b_i \left| (1-\rho^{00}) \right| b_j \right\rangle \, ,
    \nonumber \\
{\mathcal{S}'}^X_{\bar{\imath}\bar{\jmath}}
&=& +\left\langle b_{\bar{\imath}} \left| \rho^{00} \right| b_{\bar{\jmath}} \right\rangle^* \, , \nonumber \\
\mathcal{T}'_{i \bar \jmath}
&=& -\left\langle b_i \left| \rho^{00} \right| b_{\bar \jmath} \right\rangle \, .
\end{eqnarray}
The unitary transformation that diagonalizes $S'$ constitutes the second step of the procedure. As for $S$, the
transformation is block diagonal, in such a way that hole and particle states transform among themselves.
This leads to a new set of single-particle (eigen)states denoted by $| \tilde{b}_i \rangle$ and $|
\tilde{b}_{\bar{\imath}} \rangle$. The corresponding eigenvalues of $S'$ are given by
${\lambda'}^x_{i} $ and $ {\lambda'}^{b}_{\bar{\imath}}$. Again, we can classify single-particle states in three categories according to the
values of ${\lambda'}^b_{i}$ for hole states and to the values of ${\lambda'}^{x}_{\bar{\imath}}$ for particle states.

From a practical point of view, no numerical diagonalization of $S'$ is actually needed. Starting from the
properties of the states $| \tilde a_i \rangle$ and $\left| \tilde a_{\bar{\jmath}} \right\rangle$ obtained
through the diagonalization of $S$
\begin{eqnarray}
\left\langle \tilde a_i \left|(1-\rho^{11})\right|\tilde a_j \right\rangle^*
&=& \lambda^x_{i} \, \delta_{ij}
    \, , \nonumber \\
\left\langle \tilde a_{\bar{\imath}} \left| \rho^{11} \right| \tilde a_{\bar{\jmath}} \right\rangle
& = & \lambda^b_{\bar{\jmath}} \, \delta_{\bar{\imath} \bar{\jmath}} \, ,
\end{eqnarray}
one can deduce the hole and particle states of $| \Phi_1 \rangle$ that diagonalize $\mathcal{S}'^X$ and $\mathcal{S}'^B$.
We consider here the case where the eigenvalues ${\lambda }^x_{i}$ and ${\lambda}^b_{\bar{\imath}}$ differ
from zero\footnote{If the eigenvalue ${\lambda }^x_{i}/{\lambda}^b_{\bar{\imath}}$ is 0 or 1, the state $| \tilde
a_i \rangle/| \tilde a_{\bar{\imath}} \rangle$ is already a particle or a hole state of $\left| \Phi_1
\right\rangle$. These cases can be treated through a preliminary step where some of the states $| \tilde b_i
\rangle$ and $| \tilde b_{\bar{\imath}} \rangle$ are directly identified as particle and hole states of $\left|
\Phi_0 \right\rangle$. The other particle and hole states of $\left| \Phi_1 \right\rangle$ are changed accordingly
to fulfill orthogonality conditions. }. With each hole state $| \tilde a_i \rangle$, resp. particle state $|
\tilde a^+_{\bar{\imath}} \rangle$, we associate a new particle state $| \tilde b_{\bar{\imath}} \rangle$, resp.
hole state $| \tilde b_{i} \rangle$, of $\left| \Phi_1 \right\rangle$ through
\begin{eqnarray}
| \tilde b_{\bar{\imath}} \rangle
&\equiv & \frac{1}{\sqrt{\lambda^x_{i}}} \sum_{\bar k=1}^{N} \left| b_{\bar k} \right\rangle
          \langle b_{\bar k} | \tilde a_i \rangle = \frac{1-\rho^{11}}{\sqrt{\lambda^x_{i}}} \, | \tilde a_i \rangle  \, ,
\label{eq:phtilde1} \\
| \tilde b_{i} \rangle
&\equiv& \frac{1}{\sqrt{\lambda^b_{\bar{\imath}}}} \sum_{k=1}^{N} \left| b_{k} \right\rangle
         \langle b_{k} | \tilde a_{\bar{\imath}} \rangle
= \frac{\rho^{11}}{\sqrt{\lambda^b_{\bar{\imath}}}} \, | \tilde a_{\bar{\imath}} \rangle \, . \label{eq:phtilde2}
\end{eqnarray}
Those states are
\begin{itemize}
\item
orthonormal and have overlaps with the states diagonalizing $S$ given by
\begin{eqnarray}
\langle  \tilde{a}_{j} | \tilde{b}_{\bar{\imath}} \rangle
& = & +\left({\lambda}^x_{i}\right)^{1/2} \, \delta_{i j} \, ,
      \nonumber \\
\langle \tilde{a}_{\bar \jmath} | \tilde{b}_{\bar{\imath}} \rangle
& = & +\left(\lambda^x_{i}\right)^{-1/2} \, \tilde{\mathcal{T}}_{i \bar \jmath} \, ,
      \nonumber \\
\langle \tilde{a}_{\bar{\jmath}} | \tilde{b}_{i} \rangle
& = & +\left({ \lambda}^b_{\bar{\jmath}}\right)^{1/2} \, \delta_{\bar{\imath} \bar{\jmath}} \, ,
     \nonumber \\
\langle \tilde{a}_{j} | \tilde{b}_{i} \rangle
& = & -\left(\lambda^b_{\bar{\jmath}}\right)^{-1/2} \, \tilde{\mathcal{T}}_{j \bar \imath} \, ,
\end{eqnarray}
\item
eigenstates of $\mathcal{S}'^{B}$ and $\mathcal{S}'^{X}$ and fulfill
\begin{eqnarray}
\langle \tilde{b}_i | (1-\rho^{00}) | \tilde{b}_j \rangle
& = & \lambda^b_{\bar{\imath}} \, \delta_{ij} \, ,
      \nonumber \\
\langle \tilde b_{\bar{\imath}} | \rho^{00} | \tilde b_{\bar{\jmath}} \rangle^*
& = & \lambda^x_{i} \, \delta_{\bar{\imath} \bar{\jmath}}\nonumber \, ,
      \nonumber \\
\tilde{\mathcal{T}}'_{j \bar{\imath}}
& = & \widetilde{\mathcal{T}}_{i \bar{\jmath}} \, .
\end{eqnarray}
From the above relation, we deduce in  particular that ${\lambda}'^b_{i} = {\lambda}^b_{\bar{\imath}}$ and
${\lambda}'^x_{\bar{\jmath}} = {\lambda}^x_{j}$.
\end{itemize}
Now, a hole state $| \tilde b_i \rangle$  with eigenvalues ${\lambda}'^b_{ i}$ can now be decomposed as
\begin{equation}
| \tilde b_i \rangle
= | \tilde a_{\bar{\imath} } \rangle \langle \tilde a_{\bar{\imath}} |  \tilde b_i \rangle
  + \sum_{j / {\lambda}^x_{j}
= \lambda'^b_{ i}} | \tilde a_{ j } \rangle \langle \tilde a_{j} |  \tilde b_i \rangle
 \, ,
\end{equation}
where only the particle state $| \tilde{a}_{\bar{\imath} } \rangle$ from which $| \tilde{b}_i \rangle$ is
constructed appears, while the summation over holes of $| \Phi_0 \rangle$ is restricted to states that have the
same eigenvalue as $| \tilde{b}_i \rangle$. As a result, the number of states in the expansion is greatly reduced
as compared to Eq.~(\ref{eq:phoriginal}).

The matrix $A$ is not yet written in a canonical form. To complete the BMZ decomposition, we perform a third
transformation consisting of mixing the states $| \tilde{a}_i \rangle$ (resp. $| \tilde a_{\bar{\imath}} \rangle$)
within each degenerate subspace of $\mathcal{S}^X$ (resp. $\mathcal{S}^B$) in such a way that the matrix $\mathcal{A}$
(resp. $\mathcal{Y}$) becomes diagonal. The product of the transformations diagonalizing $S$ and $A$ provides the
unitary transformation $D$ introduced in Sec.~\ref{illdefined} where we explained our method in the general
case. As $A$ is being diagonalized, the states $| \tilde b_{\bar{\imath}} \rangle$ (resp. $| \tilde b_{i}
\rangle$) are modified accordingly through Eq.~(\ref{eq:phtilde1}) (resp. Eq.~(\ref{eq:phtilde2}))\footnote{Although
the last step bringing $A$ into a canonical form involves an additional transformation, we use the same notation
$\{| \tilde a_{i} \rangle ; | \tilde a_{\bar{\imath}} \rangle\}_{i=1, \ldots, N}$ and $\{| \tilde b_{i} \rangle ;
| \tilde b_{\bar{\imath}} \rangle\}_{i=1,\ldots, N}$ to denote the bases before and after that last
transformation.}. Together with the transformation that diagonalized $S'$, this provides the unitary
transformation $C$ introduced in Sec.~\ref{illdefined}.

\begin{figure}[t!]
\includegraphics[height=8.cm,angle=-90]{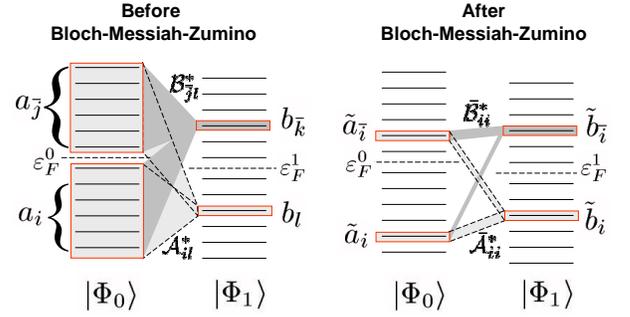}
\caption{(Color online) Schematic illustration of the different bases introduced in the text. Left: the Slater determinants $| \Phi_{0} \rangle$ and $| \Phi_{1} \rangle$ are expressed in their respective natural bases $\{| a_{i} \rangle\}$ and $\{| b_{l} \rangle\}$. The shaded areas indicate that one state of the basis $\{| b_{l} \rangle\}$ ($\{| a_{i} \rangle\}$) is a priori spread over all particle and hole states of the basis $\{| a_{i} \rangle\}$
($\{| b_{l} \rangle\}$). Right: the BMZ decomposition leads to two bases which are such that each transformed state $|
\tilde{b}_{i} \rangle$ ($| \tilde{a}_{i} \rangle$) decomposes only onto two states $(\tilde a_{i}, \tilde
a_{\bar{\imath}})$ ($(\tilde b_{i}, \tilde b_{\bar{\imath}})$) (one hole and one particle). At the same time, the
latter decomposition highlights that the notion of \emph{conjugated} pairs for Slater determinants relates to the
association of each hole state with a particular particle state. Note that energy levels have no specific meaning
in this figure; they are just meant to characterize occupied and empty levels for each of the Slater
determinants.} \label{fig:pole1}
\end{figure}

Finally, the BMZ decomposition of the Bogoliubov transformation between two Slater determinants
(\ref{eq:bogo:slater}) is achieved in the sense that the matrices
\begin{equation}
\bar{A}
\equiv \left( \begin{array}{cc}
       \bar{\mathcal{A}} & 0 \\
       0 & \bar{\mathcal{Y}}
       \end{array} \right)_{2N} \, ,
\qquad
\bar{B}
\equiv \left( \begin{array}{cc}
       0 & \bar{\mathcal{X}} \\
       \bar{\mathcal{B}}  & 0
       \end{array} \right)_{2N} \, ,
\end{equation}
are in canonical forms when expressed in the new bases.  In the present case, conjugated pairs are made of a
particle state $\tilde a^+_{\bar{\imath}}$ and a hole state $\tilde a^+_i$ of $| \Phi_0 \rangle$. Through a
BCS-like transformation, the latter pair is associated with a unique particle-hole pair $(\tilde b^+_i , \tilde
b^+_{\bar{\imath}})$ of $\left| \Phi_1 \right\rangle$\footnote{The bases $\{\tilde a_i, \tilde a^{\dagger}_i\}$
and $\{\tilde b_i, \tilde b^{\dagger}_i\}$ are the analog of the bases $\{\tilde \alpha_\nu, \tilde
\alpha^{\dagger}_\nu\}$ and $\{\tilde \beta_\mu, \tilde \beta^{\dagger}_\mu\}$ defined by Eq.~(\ref{eq:bcslike}) in
the general case.}
\begin{eqnarray}
\tilde b_i
& = &  \bar{\mathcal{A}}_{i i} \, \tilde a_i + \bar{\mathcal{B}}_{\bar{\imath} i} \, \tilde a_{\bar{\imath}} \,, \\
\tilde b^+_{\bar{\imath}}
& = &  \bar{\mathcal{X}}_{i {\bar{\imath}}} \, \tilde a^+_i + \bar{\mathcal{Y}}_{\bar{\imath} \bar{\imath}} \, a^+_{\bar{\imath}} \, ,
\end{eqnarray}
where the matrix elements of $\bar{A}$ and $\bar{B}$ are given by the canonical expression
\begin{alignat}{3}
\label{eq:uvxybar} \bar{\mathcal{A}}_{ji}
& \equiv  \delta_{ji} \langle \tilde b_i | \tilde a_i \rangle \, , & \qquad \bar{\mathcal{B}}_{\bar{\jmath} i}
& \equiv \delta_{ \bar{\jmath} \bar{\imath}} \langle \tilde  b_i | \tilde a_{\bar{\imath}} \rangle \, ,
         \nonumber \\
\bar{\mathcal{X}}_{j \bar{\imath}}
& \equiv \delta_{ \bar{\jmath} \bar{\imath}}  \langle   \tilde a_i  | \tilde b_{\bar{\imath}} \rangle \, ,
& \bar{\mathcal{Y}}_{\bar{\jmath} \bar{\imath}}
& \equiv \delta_{ \bar{\jmath} \bar{\imath}} \langle \tilde a_{\bar{\imath}}  | \tilde b_{\bar{\imath}} \rangle \, .
\end{alignat}
Finally, Fig.~\ref{fig:pole1} illustrates the procedure and the meaning of putting the Bogoliubov transformation
of dimension $2N$ that connects two Slater determinants into a canonical form.

\subsection{Correction for spurious processes}

Given the BMZ decomposition of the transformation between two Slater determinants obtained above, we can apply
directly the results derived in Sec.~\ref{illdefined}, using the new basis $\{\tilde a_{i}, \tilde
a^{\dagger}_{i}\}$, to write the state $\left| \Phi_1 \right\rangle$ under the form
\begin{equation}
\label{stateSD}
\left| \Phi_1 \right\rangle
= \prod_{i=1}^{N} \left(  \bar{\mathcal{A}}^*_{i i}
                        + \bar{\mathcal{B}}^*_{i \bar{\imath} } \; \tilde{a}_i \, \tilde{a}^+_{\bar{\imath}}
                  \right) | \Phi_0 \rangle \, ,
\end{equation}
and the transition density matrix between the two Slater determinants as
\begin{equation}
\label{denstransSD}
\rho^{01}
= \rho^{00}
  + \sum_{i=1}^{N} | \tilde a_{\bar{\imath}} \rangle \, \bar{\mathcal{Z}}_{\bar{\imath} i} \, \langle \tilde a_i |
\, ,
\end{equation}
where, as before, $\bar{\mathcal{Z}}_{\bar{\imath} i} \equiv \bar{\mathcal{B}}^*_{\bar{\imath} i} \; \bar{\mathcal{A}}^{*-1}_{ii}$.
As can be seen from Eq.~(\ref{stateSD}), the picture emerging in the new basis is very intuitive
since the Slater determinant $\left| \Phi_1 \right\rangle$ is obtained as a linear combination of up to, very
specific, N particle-N hole configurations on the vacuum $\left| \Phi_0 \right\rangle$. In particular, each hole
$\tilde a_{i}$ is combined with a specific particle $\tilde a_{i}$; the latter association defining the notion of
\emph{conjugated pair} for the Bogoliubov transformation that links two Slater determinants. Such an expression
simplifies tremendously the general expression connecting two non-orthogonal Slater determinants where each hole
state can be excited into any particle states, up to infinite energy.

As for calculations with an explicit treatment of pairing, Eq.~(\ref{denstransSD}) can be used to express the MR
energy kernel on the basis of the GWT. This leads to
\begin{eqnarray}
\mathcal{E}_{GWT}[0,1]
&=& \sum_{i=1}^{N} \Big( t_{\tilde a_i \tilde a_i}
+ t_{\tilde a_i \tilde a_{\bar{\imath}}} \, \bar{\mathcal{Z}}_{ \bar{\imath} i} \Big) \nonumber \\
&+& \frac{1}{2} \sum_{i,j=1}^{N} \bar{v}^{\rho\rho}_{\tilde a_i \tilde a_j \tilde a_i \tilde a_j
} \nonumber \\
&+& \frac{1}{2} \sum_{i,j=1}^{N} \Big(\bar{v}^{\rho\rho}_{\tilde a_i \tilde a_j \tilde a_{\bar{\imath}} \tilde a_j}
    + \bar{v}^{\rho\rho}_{\tilde a_j \tilde a_i \tilde a_j \tilde a_{\bar{\imath}}} \Big)
    \, \bar{\mathcal{Z}}_{ \bar{\imath} i} \nonumber \\
&+& \frac{1}{2} \sum_{i,j=1}^{N} \bar{v}^{\rho\rho}_{\tilde a_i \tilde a_j \tilde a_{\bar{\imath}} \tilde
a_{\bar{\jmath}}} \, \bar{\mathcal{Z}}_{\bar{\imath} i } \, \bar{\mathcal{Z}}_{\bar{\jmath} j} \, .
\end{eqnarray}
Starting from this expression, the self-interaction $\mathcal{E}^{\rho\rho}_{SI}[0,1]$ and the spurious contribution
to the MR energy due to the construction of the energy kernels on the basis of the GWT
$\mathcal{E}^{\rho\rho}_{CG}[0,1]$ can be identified
\begin{eqnarray}
\mathcal{E}^{\rho\rho}_{SI}[0,1] & =& \frac{1}{2} \sum_{i=1}^{N}
\bar{v}^{\rho\rho}_{\tilde a_i \tilde a_i \tilde a_i \tilde a_i} \nonumber \\
& +& \frac{1}{2} \sum_{i=1}^{N} \Big(\bar{v}^{\rho\rho}_{\tilde a_i \tilde a_i \tilde a_{\bar{\imath}} \tilde a_i}
+ \bar{v}^{\rho\rho}_{\tilde a_i \tilde a_i \tilde a_i \tilde a_{\bar{\imath}}} \Big)
\, \bar{\mathcal{Z}}_{ \bar{\imath} i} \, , \\
\mathcal{E}^{\rho\rho}_{CG}[0,1] &=& \frac{1}{2} \sum_{i=1}^{N} \bar{v}^{\rho\rho}_{\tilde a_i \tilde a_i \tilde
a_{\bar{\imath}} \tilde a_{\bar{\imath}}} \, \bar{\mathcal{Z}}^2_{\bar{\imath} i } \, , \label{CGHF}
\end{eqnarray}
whereas, of course, no spurious self-pairing occurs in the present application.

Again, both terms are zero for energy kernels obtained as the matrix element of a Hamiltonian. The latter of the
two terms is very likely to be responsible for the difficulties recently encountered in calculations aiming at
restoring angular momentum using a MR set of cranked Slater determinants~\cite{doba06a}. We thus advocate the
removal of the contribution given by Eq.~(\ref{CGHF}) in such a context.

\section{Conclusions}
\label{conclusions}

The present work concentrates on multi-reference (MR) calculations, customarily called "beyond-mean-field"
calculations in the literature, performed within the Energy Density Functional (EDF) formalism. The
multi-reference method is nowadays routinely applied with the aim of including long-range correlations associated
with large-amplitude collective motions which are difficult to incorporate in a more traditional single-reference
(SR), i.e.\ "mean-field", EDF formalism~\cite{bender03b}. So far, the framework for such MR-EDF calculations was
set-up by analogy with projection techniques and the Generator Coordinate Method (GCM), that has been rigorously
formulated so far only within a Hamiltonian/wave-function-based framework~\cite{Won75a,ring80a}.

The first achievement of the present work is to demonstrate that the usual extension of the single-reference
energy functional $\mathcal{E}[\rho,\kappa,\kappa^{\ast}]$ into the non-diagonal energy kernel $\mathcal{E}[0,1]$ at
play in MR calculations through $\mathcal{E}[0,1]\equiv\mathcal{E}[\rho^{01},\kappa^{01},\kappa^{10 \, \ast}]$ is ill
defined. The latter extension, based on the Generalized Wick Theorem (GWT)~\cite{balian69a} is well defined within
a Hamiltonian/wave-function-based projected-GCM framework, but happens to be at the origin of spurious
divergences and steps in MR-EDF calculations, as recently realized for particle number and angular momentum
restorations~\cite{anguiano01b,doba05a,doba06a}.

The second achievement of the present paper is to propose a method to identify, for any type of symmetry
restoration and/or GCM-based calculations, the spurious terms in the MR-EDF responsible for divergences, steps,
self-interaction and self-pairing. The versatility of the method also allows to take care of difficulties
encountered when mixing within the MR-EDF calculation states obtained from quasi-particle
excitations~\cite{tajima92a}.

In practice, the method requires first to put the Bogoliubov transformation connecting two vacua $| \Phi_0
\rangle$ and $| \Phi_1 \rangle$ into a canonical form. In the corresponding canonical basis, spurious contributions to the
MR energy kernel $\mathcal{E}_{GWT}[0,1]$ can be identified and, if necessary, removed. Subtracting the terms
responsible for divergences and steps is a prerequisite to perform well-defined multi-reference EDF calculations,
and that is what we advocate in the present work. However, the quantitative impact of self-interaction and
self-pairing processes on observable as well as the technical difficulties associated with their removal remain to
be studied. As they do not lead to dramatic consequences, i.e.\ divergences and steps, this can be postponed.

As the proposed procedure must be implemented for all pairs of vacua involved in the multi-reference calculation,
the correction increases the basic computation cost. On the other hand, the number of summations in the
calculation of the MR energy is significantly reduced in the canonical basis extracted to remove the spurious contributions.
Thus, it might be of interest to take advantage of that fact in actual implementations of the MR-EDF formalism.
This is the goal of a forthcoming publication to discuss such an implementation. In the meantime, another
publication is devoted to illustrating the features of the correction method proposed in the present paper by
applying it to the rather transparent case of Particle Number Restoration~\cite{bender07x}.

Despite the success of regularized MR-EDF calculations documented in Ref.~\cite{bender07x}, some important questions remain
unanswered as of today regarding the foundation and the implementation of MR methods within a consistent EDF
framework. First, it is unclear in which sense the "projected" functional that one manipulates in such
calculations can be attributed, at least implicitly, to a state belonging to a specific irreducible representation
of the symmetry group of the nuclear Hamiltonian. The latter statement may seem rather paradoxical as the main
purpose of those calculations is precisely to "restore the symmetries". The crucial point is that, as opposed to
the \emph{strict} Projected Hartree-Fock-Bogoliubov approach~\cite{Won75a,Man75a,ring80a}, the MR energy functional
is not obtained as the average value of a Hamiltonian in the projected state but as a functional of
(non-observable) transition densities defined for each pair of vacua belonging to the MR set. The rather
complicated dependence on those densities and the use of different effective vertices (interactions) in different
channels of the functional forbid the explicit re-factorization of the energy in terms of the projected state. The
benefit of such a method is that correlations which go beyond the strict projected-GCM approximation can be easily
incorporated into the MR formalism. Again, the drawback is that it is not clear in which sense symmetries are
actually restored in existing MR-EDF calculations. The clarification of such a question and the proper formulation
of symmetry restoration within a well-defined MR-EDF theory is mandatory in the near future.

Last but not least, the present work raises three important questions; the first two being related to the
construction of energy density functionals; i.e.\ Skyrme, Gogny or any other types of realistic functionals.
\begin{itemize}
\item[(i)]
Only the removal of the most dangerous spurious terms associated with the use of the GWT as a basis to construct
phenomenological MR energy kernels has been advocated in the present paper and implemented in
Ref.~\cite{bender07x}. However, it is worth noticing that SR- and MR-EDF calculations are also plagued with less
dangerous pathologies related to \emph{self-interaction}~\cite{perdew81a} and \emph{self-pairing}~\cite{bender07x}. In particular, such pathologies contaminate MR energy kernels independently on whether the SWT or the GWT is used as a basis to define them.
At this point though, we do not advocate the correct for self-interaction and self-pairing
processes briefly discussed in the present paper. The reason is that such additional corrections will impact
explicitly the SR functional on which the MR calculation is based. This means that self-interaction- and
self-pairing-free SR functionals must be constructed in order to test the importance of such spurious processes.
This is a non-trivial step to take because doing so will modify the structure of usual functionals and most
importantly the way self-consistent SR calculations usually work~\cite{perdew81a}. Such a work is underway.
\item[(ii)]
The correction method proposed in the present work is based on using the Hamiltonian/wave-function framework,
together with the Standard Wick Theorem~\cite{wick50a}, as a reference point. A non trivial implication is that
the procedure provides a way to correct functionals which depends only on \emph{integer} powers of the normal and
anomalous density matrices~\cite{duguet07a}. As we do not see at this point how to proceed otherwise, the present
work acts as a strong motivation to construct SR functionals that only involve integer powers of the density
matrices in the near future. For reasons that are well known to practitioners, functionals depending on the third power in the local density do not
work well enough and it remains to be seen whether or not using fourth powers of the local density is sufficient
to construct high-precision energy density functionals.
\item[(iii)] The present work is a
satisfactory solution to an extreme problem faced by MR energy density functional calculations. However, it is not
entirely satisfying from a fundamental point of view since it amounts to correcting a phenomenological
construction of multi-reference energy kernels that is ill-defined in the first place. Combined with related
issues regarding restoration of symmetries, the present work calls for formal developments towards the derivation
of a MR-EDF formalism from first principles. In particular, the regularization of the energy kernels happens to be basis-dependent, just as standard self-interaction correction methods in DFT are~\cite{perdew81a}. A nice feature of the practical method we use to proceed to the regularization is that it leads to a unique basis among all possible ones. It is so because, given $| \Phi_0 \rangle$ and $| \Phi_1 \rangle$, the application of the BMZ decomposition defines a unique basis, independently on the actual representation of the two states. Still, the fact that the removal of spurious contributions is basis-dependent in the first place is not satisfactory and deserves some further attention in the future.
\end{itemize}

%
%

\begin{acknowledgments}
The authors thank M.~Stoitsov, W.~Nazarewicz, J.~Dobaczewski and P.-G.~Reinhard for many enlightening discussions,
and particularly for the communication of their results prior to publication, which inspired this study. This work
was supported by the U.S. National Science Foundation under Grant No.\ PHY-0456903. M.~B.\ thanks the NSCL for the
kind hospitality  during the completion of this work. The authors also thank T. Lesinski and K. Bennaceur for the
critical reading of the manuscript.
\end{acknowledgments}

%
%

\appendix

\section{Expression of $\langle \Phi_0 | v_{123} | \Phi_1 \rangle$ }
\label{sec:threebody}

Using the method depicted in Sec.~\ref{sec:higherorder}, we
obtained the following expression:
\begin{widetext}
\begin{eqnarray}
\langle \Phi_0 | \hat v_{123} | \Phi_1 \rangle&=&
~~\frac{1}{6} \sum  \tilde V^{0*}_{l \nu} \tilde V^0_{i\nu} \tilde
V^{0*}_{m \mu} \tilde V^0_{j\mu}
 V^{0*}_{n \lambda} \tilde V^0_{k\lambda}
\, \bar{v}_{ijklmn} \, \langle \Phi_0   | \Phi_1 \rangle \nonumber \\
&&+\frac{1}{6} \sum  \tilde V^{0*}_{l \nu} \tilde V^0_{i\nu}
\tilde V^{0*}_{m \mu} \tilde V^0_{j\mu} \tilde U^0_{n \bar
\lambda} \tilde V^0_{k\lambda} \, \bar{v}_{ijklmn} \, \bar
B^{*}_{\bar \lambda \lambda}
\langle \Phi_0  | \Phi_1, \lambda \rangle \nonumber \\
&&+ \frac{1}{6} \sum  \tilde V^{0*}_{l \nu} \tilde V^0_{i\nu}
\tilde U^0_{m \bar \mu} \tilde V^0_{j\mu} V^{0*}_{n \lambda}
\tilde V^0_{k\lambda} \, \bar{v}_{ijklmn} \,
\bar B^*_{\bar \mu \mu} \langle \Phi_0  | \Phi_1, \mu \rangle \nonumber \\
&&+ \frac{1}{6} \sum   \tilde U^0_{l \bar \nu} \tilde V^0_{i\nu}
\tilde V^{0*}_{m \mu} \tilde V^0_{j\mu} V^{0*}_{n \lambda} \tilde
V^0_{k\lambda} \, \bar{v}_{ijklmn} \, \bar B^*_{\bar \nu \nu}
\langle \Phi_0  | \Phi_1, \nu \rangle \nonumber \\
&&+ \frac{1}{6} \sum  \tilde U^0_{l \bar \nu} \tilde V^0_{i\nu}
\tilde U^0_{m \bar \mu} \tilde V^0_{j\mu} V^{0*}_{n \lambda}
\tilde V^0_{k\lambda} \, \bar{v}_{ijklmn} \, \bar B^*_{\bar \nu
\nu} \bar B^*_{\bar \mu \mu}
\langle \Phi_0  | \Phi_1, \nu, \mu \rangle \nonumber \\
&&+ \frac{1}{6} \sum  \tilde V^{0*}_{l \nu} \tilde V^0_{i\nu}
\tilde U^0_{m \bar \mu} \tilde V^0_{j\mu} \tilde U^0_{n \bar
\lambda} \tilde V^0_{k\lambda} \, \bar{v}_{ijklmn} \, \bar
B^*_{\bar \lambda \lambda} \bar B^*_{\bar \mu \mu}
\langle \Phi_0  | \Phi_1, \mu, \lambda \rangle \nonumber \\
&&+ \frac{1}{6} \sum  \tilde U^0_{l \bar \nu} \tilde V^0_{i\nu}
\tilde V^{0*}_{m \mu} \tilde V^0_{j\mu}
 \tilde U^0_{n \bar \lambda} \tilde V^0_{k\lambda} \, \bar{v}_{ijklmn} \,
\bar B^*_{\bar \nu \nu}\bar B^*_{\bar \lambda \lambda}
\langle \Phi_0  | \Phi_1, \nu, \lambda \rangle \nonumber \\
&&+ \frac{1}{6} \sum  \tilde U^0_{l \bar \nu} \tilde V^0_{i\nu}
\tilde U^0_{m \bar \mu} \tilde V^0_{j\mu} \tilde U^0_{n \bar
\lambda} \tilde V^0_{k\lambda} \, \bar{v}_{ijklmn} \, \bar
B^*_{\bar \nu \nu} \bar B^*_{\bar \mu \mu}  \bar B^*_{\bar \lambda
\lambda}
\langle \Phi_0  | \Phi_1, \nu, \mu, \lambda \rangle \nonumber \\
&&+\frac{1}{4} \sum \tilde V^{0*}_{l \nu} \tilde V^0_{i\nu} \tilde
V^0_{j \mu} \tilde U^{0*}_{k \mu} \tilde V^{0*}_{m \lambda} \tilde
U^0_{n\lambda} \, \bar{v}_{ijklmn} \,
\langle \Phi_0  | \Phi_1 \rangle  \nonumber \\
&&+ \frac{1}{4} \sum \tilde V^{0*}_{l \nu} \tilde V^0_{i\nu}
\tilde V^0_{j \mu} \tilde U^{0*}_{k \mu} \tilde U^0_{m \bar
\lambda} \tilde U^0_{n \lambda} \, \bar{v}_{ijklmn} \,  \bar
B^*_{\bar \lambda \lambda}
\langle \Phi_0  | \Phi_1, \lambda \rangle \nonumber \\
&&+ \frac{1}{4} \sum \tilde V^{0*}_{l \nu} \tilde V^0_{i\nu}
\tilde V^0_{j \mu}  \tilde V^0_{k \bar \mu} \tilde V^{0*}_{m
\lambda} \tilde U^0_{n\lambda} \, \bar{v}_{ijklmn} \,
\bar B^*_{ \bar \mu \mu} \langle \Phi_0  | \Phi_1, \mu \rangle \nonumber \\
&&+  \frac{1}{4} \sum  \tilde U^0_{l \bar \nu} \tilde V^0_{i\nu}
 \tilde V^0_{j \mu} \tilde U^{0*}_{k \mu}
\tilde V^{0*}_{m \lambda} \tilde U^0_{n\lambda} \,
\bar{v}_{ijklmn} \, \bar B^*_{\bar \nu \nu}
\langle \Phi_0  | \Phi_1, \nu \rangle \nonumber \\
&&+ \frac{1}{4}  \sum \tilde V^{0*}_{l \nu} \tilde V^0_{i\nu}
\tilde V^0_{j \mu}  \tilde V^0_{k \bar \mu} \tilde U^0_{m \bar
\lambda} \tilde U^0_{n \lambda} \, \bar{v}_{ijklmn} \, \bar B^*_{
\bar \mu \mu}\bar B^*_{\bar \lambda \lambda}
\langle \Phi_0  | \Phi_1, \mu, \lambda \rangle \nonumber \\
&&+\frac{1}{4}  \sum \tilde U^0_{l \bar \nu} \tilde V^0_{i\nu}
\tilde V^0_{j \mu} \tilde U^{0*}_{k \mu} \tilde U^0_{m \bar
\lambda} \tilde U^0_{n \lambda} \, \bar{v}_{ijklmn} \, \bar
B^*_{\bar \nu \nu}\bar B^*_{\bar \lambda \lambda}
\langle \Phi_0  | \Phi_1, \nu, \lambda \rangle \nonumber \\
&&+\frac{1}{4}  \sum \tilde U^0_{l \bar \nu} \tilde V^0_{i\nu}
\tilde V^0_{j \mu}  \tilde V^0_{k \bar \mu} \tilde V^{0*}_{m
\lambda} \tilde U^0_{n\lambda} \, \bar{v}_{ijklmn} \, \bar
B^*_{\bar \nu \nu}\bar B^*_{ \bar \mu \mu}
\langle \Phi_0  | \Phi_1, \nu, \mu \rangle \nonumber \\
&&+\frac{1}{4}  \sum \tilde U^0_{l \bar \nu} \tilde V^0_{i\nu}
\tilde V^0_{j \mu}  \tilde V^0_{k \bar \mu} \tilde U^0_{m \bar
\lambda} \tilde U^0_{n \lambda} \, \bar{v}_{ijklmn} \, \bar
B^*_{\bar \nu \nu}\bar B^*_{ \bar \mu \mu} \bar B^*_{\bar \lambda
\lambda} \langle \Phi_0  | \Phi_1, \nu, \mu, \lambda
\rangle \label{eq:threedirect}
\end{eqnarray}
\end{widetext}
where the sums run over all the indices, with the convention that when two or more $\nu_i$'s belong to the same
conjugated pair, $\langle \Phi_0  | \Phi_1, \nu_1, \cdots,\nu_n \rangle = 0$. In the latter expression, the first 8 terms come from the expansion of $\rho^{01}\rho^{01}\rho^{01}$. Among these terms, the four last one will provide spurious contributions to
$\mathcal{E}^{\rho\rho\rho}_{GWT}$. Similarly, the remaining 8 terms originate from $\rho^{01} \kappa^{10^*} \kappa^{01}$
whereas the last 4 of them provide spurious contribution to $\mathcal{E}^{\rho\kappa\kappa}_{GWT}$.

\bibliography{papierI}

\end{document}